\pdfoutput=1
\documentclass[sigconf,]{acmart}

\makeatletter
\def\@ACM@copyright@check@cc{}
\makeatother

\copyrightyear{2025}
\acmYear{2025}
\setcopyright{cc}
\setcctype[4.0]{by}
\acmConference[SIGIR '25]{Proceedings of the 48th International ACM SIGIR Conference on Research and Development in Information Retrieval}{July 13--18, 2025}{Padua, Italy}
\acmBooktitle{Proceedings of the 48th International ACM SIGIR Conference on Research and Development in Information Retrieval (SIGIR '25), July 13--18, 2025, Padua, Italy}\acmDOI{10.1145/3726302.3729983}
\acmISBN{979-8-4007-1592-1/2025/07}

\makeatother

\usepackage{enumitem}
\newlist{inlinelist}{enumerate*}{1}
\setlist*[inlinelist,1]{%
  label=(\roman*),
}
\usepackage{subfigure}
\usepackage{algorithm}
\usepackage[noend]{algpseudocode}
\usepackage{multirow} 
\usepackage{rotating}
\usepackage{tikz}
\usepackage{amsmath}
\usepackage{lipsum,adjustbox}
\usepackage{vcell}
\usepackage{xcolor}
\usepackage{xspace}
\usepackage{diagbox}
\usepackage{pifont}
\usepackage{enumitem}
\usepackage{balance}
\usepackage{cleveref}
\usepackage{placeins}
\usepackage{graphicx}
\usepackage{pgfplots}
\usepackage[normalem]{ulem}
\usetikzlibrary{shapes,arrows,positioning,fit,calc,matrix}
\usepgfplotslibrary{groupplots}

\newcommand{\TasBSymbol}{\spadesuit}
\newcommand{\BMSymbol}{\dagger}
\newcommand{\CLDRDSymbol}{\diamondsuit}
\newcommand{\BeBaseSymbol}{\clubsuit}

\title{Hypencoder: Hypernetworks for Information Retrieval}

\author{Julian Killingback}
\affiliation{%
  \institution{University of Massachusetts Amherst}
  \city{Amherst}
  \state{MA}
  \country{United States}
}
\email{jkillingback@cs.umass.edu}

\author{Hansi Zeng}
\affiliation{%
  \institution{University of Massachusetts Amherst}
  \city{Amherst}
  \state{MA}
  \country{United States}
}
\email{hzeng@cs.umass.edu}

\author{Hamed Zamani}
\affiliation{%
  \institution{University of Massachusetts Amherst}
  \city{Amherst}
  \state{MA}
  \country{United States}
}
\email{zamani@cs.umass.edu}

\pgfplotsset{compat=1.18}
\begin{document}
\newcommand{\name}{Hypencoder}
\newcommand{\mininame}{$q$-net}
\newcommand{\hyperheadlayer}{hyperhead layer}
\newcommand{\todo}[1]{\textcolor{red!70}{TODO: #1}}
\newcommand{\ItemEncoderName}{$\Psi$}
\newcommand{\ItemRepsName}{$E_d$}
\newcommand{\QueryEncoderName}{$\Phi$}
\newcommand{\HyperHeadName}{hyper-head}

\begin{abstract}
Existing information retrieval systems are largely constrained by their reliance on vector inner products to assess query-document relevance, which naturally limits the expressiveness of the relevance score they can produce. We propose a new paradigm; instead of representing a query as a vector, we use a small neural network that acts as a learned query-specific relevance function. This small neural network takes a document representation as input (in this work we use a single vector) and produces a scalar relevance score. To produce the small neural network we use a hypernetwork, a network that produces the weights of other networks, as our query encoder. We name this category of encoder models Hypencoders. Experiments on in-domain search tasks show that Hypencoders significantly outperform strong dense retrieval models and even surpass reranking models and retrieval models with an order of magnitude more parameters. To assess the extent of Hypencoders' capabilities, we evaluate on a set of hard retrieval tasks including tip-of-the-tongue and instruction-following retrieval tasks. On harder tasks, we find that the performance gap widens substantially compared to standard retrieval tasks. Furthermore, to demonstrate the practicality of our method, we implement an approximate search algorithm and show that our model is able to retrieve from a corpus of 8.8M documents in under 60 milliseconds.
\end{abstract}




\maketitle

\section{Introduction}
Efficient neural retrieval models are based on a bi-encoder (or two-tower) architecture, in which queries and documents are represented separately using either high-dimensional sparse \cite{SNRM,SPLADE, SPLADE++} or relatively low-dimensional dense vectors \cite{DPR,ANCE,ColBERT_v1, CL-DRD, CoCondensor}. These models use simple and lightweight similarity functions, e.g., inner product or cosine similarity, to compute the relevance score for a given pair of query and document representations. We demonstrate theoretically that inner product similarity functions fundamentally limit the types of relevance that retrieval models can express. Specifically, we prove that there is always a set of relevant documents which cannot be perfectly retrieved regardless of the query vector and specific encoder model.

Motivated by this theoretical argument, we introduce a new category of retrieval models that can capture complex relationships between query and document representations. Building upon the hypernetwork literature in machine learning \cite{HyperNetworks, HyperShot, ContinualLearningWithHypernetworks}, we propose \name{}--a generic framework that learns a query-dependent multi-layer neural network as a similarity function that is applied to the document representations. In more detail, \name{} applies attention-based hypernetwork layers, called \hyperheadlayer{}s, to the contextualized query embeddings output by a backbone transformer encoder. Each \hyperheadlayer{} produces the weight and bias matrices for a neural network layer in the query-dependent similarity network, called the \mininame{}. The \mininame{} is then applied to each document representation, which results in a scalar relevance score. We demonstrate that the \name{} framework can be optimized end-to-end and used for efficient retrieval from a large corpus. Specifically, we propose a graph-based greedy search algorithm that approximates exhaustive retrieval using \name{} while being substantially more efficient. 

We conduct extensive experiments on a wide range of datasets to demonstrate the efficacy of \name{}. We show that our implementation of \name{} outperforms strong single-vector dense and sparse retrieval models on MSMARCO \cite{MSMARCO} and TREC Deep Learning Track data \cite{TREC_DL_2019, TREC_DL_2020}, in addition to complex retrieval tasks, such as TREC DL-Hard \cite{TREC_DL_HARD}, TREC Tip-of-the-Tongue (TOT) Track \cite{TREC-TOT-2023}, and the instruction following dataset FollowIR \cite{FollowIR}. Across these benchmarks, \name{} shows consistent performance gains. Using the proposed approximate retrieval approach on MSMARCO \cite{MSMARCO}, with approximately 8.8 million passages, results in an average search latency of 59.6 milliseconds per query on a single NVIDIA L40S GPU.

A main advantage of hypernetworks in machine learning is their ability to learn generalizable representations. To demonstrate that \name{} also inherits this generalization quality, we evaluated our model under various domain adaptation settings: (1) adaptation to question answering datasets in biomedical and financial domains, and (2) adaptation to other retrieval tasks, including entity and argument retrieval, where \name{} again demonstrates superior performance compared to the baselines.

We believe that these performance gains are a byproduct of our main contributions: \name{} introduces a new way to think about what retrieval and relevance functions can be, it opens a new world of possibilities by bridging the gap between neural networks and retrieval similarity functions. We believe \name{} is especially important at this time given the new demands for longer and more complex queries brought on by the widespread usage of large language models and it is our belief that \name{} represents an important step towards this goal. To help facilitate this goal, we have open sourced all of our code for training, retrieval, and evaluation.\footnote{Available at \url{https://github.com/jfkback/hypencoder-paper}}

\definecolor{base_encoder_color}{HTML}{dbf0ea}
\definecolor{base_encoder_outline_color}{HTML}{43aa8b}
\definecolor{pooler_color}{HTML}{d6ecf5}
\definecolor{pooler_outline_color}{HTML}{277da1}
\definecolor{score_color}{HTML}{fdefce}
\definecolor{score_outline_color}{HTML}{f9c74f}
\definecolor{q_net_border_color}{HTML}{abc4ff}
\definecolor{q_net_fill_color}{HTML}{d7e3fc}
\definecolor{q_net_connection_color}{HTML}{a6a6a6}
\definecolor{gray}{HTML}{a6a6a6}
\definecolor{hyper_head_color}{HTML}{fde8ce}
\definecolor{hyper_head_outline_color}{HTML}{f8961e}
\definecolor{color4}{HTML}{c0fdff}
\begin{figure*}[t]
\scalebox{2.0}{
\begin{tikzpicture}
\tikzset{
  pics/encoder/.style  n args={4}{code={
    \node[rectangle, minimum width=9.708mm, minimum height=4mm, fill=base_encoder_color, draw=base_encoder_outline_color, rounded corners=0.5mm] (base_encoder) {\tiny encoder};
    \node[rectangle, minimum width=9.708mm, minimum height=2.4mm, fill=#3, draw=#4, rounded corners=0.5mm, font=\tiny] [above = 0.5mm of base_encoder] (pooler) {#2};
    \node[fill=none, outer sep=2.0] [below =1.5 mm of base_encoder] (query_input) {\small #1};
    \draw[->, line width=0.2mm] (query_input) -- (base_encoder);
    \draw[-, line width=0.2mm] (base_encoder) -- (pooler);    
  }},
}

\tikzset{
  pics/crossEncoder/.style={code={
    \node[rectangle, minimum width=9.708mm, minimum height=4mm, fill=base_encoder_color, draw=base_encoder_outline_color, rounded corners=0.5mm] (base_encoder) [right = 3.0mm of query_encoder3] {\tiny encoder};
    \node[rectangle, minimum width=9.708mm, minimum height=2.4mm, fill=pooler_color, draw=pooler_outline_color, rounded corners=0.5mm] [above = 0.5mm of base_encoder] (pooler) {\tiny score-head};
    \node[fill=none, outer sep=2.0, xshift=2.5mm] [below =1.5 mm of base_encoder] (query_input) {\tiny q};
    \node[fill=none, outer sep=2.0, xshift=-2.5mm] [below  =1.5 mm of base_encoder] (passage_input) {\tiny d};
    \draw[->, line width=0.2mm] (query_input.north) -- ++(0pt,1.5mm);
    \draw[->, line width=0.2mm] (passage_input.north) -- ++(0pt,1.5mm);
    \draw[-, line width=0.2mm] (base_encoder) -- (pooler); 
  }},
}

\tikzset{
  pics/score/.style={code={
    \node[circle, fill=score_color, draw=score_outline_color, minimum width=2mm, minimum height=2mm]{\tiny s};
  }},
}

\tikzset{
  pics/qnet/.style={code={
      \foreach \N [count=\lay,remember={\N as \Nprev (initially 0);}]
                   in {3, 3, 3, 1}{ %
        \foreach \i [evaluate={\y=\N/2-\i; \x=\lay; \prev=int(\lay-1);}]
                     in {1,...,\N}{ %
          \node[minimum size=1.0mm, circle, draw=q_net_border_color, fill=q_net_fill_color, inner sep=0pt] (N\lay-\i) at (\x / 5.5,\y /5.5) {};
          \ifnum\Nprev>0 %
            \foreach \j in {1,...,\Nprev}{ %
              \draw[line width=0.1mm, draw=q_net_connection_color] (N\prev-\j) -- (N\lay-\i);
            }
          \fi
        }
      }  
  }},
}

\node[matrix, inner sep=0] (passage_encoder1)  {\pic{encoder={\tiny d}{pooler}{pooler_color}{pooler_outline_color}};\\}; 
\node[matrix, inner sep=0] (query_encoder1) [right =1.0 mm of passage_encoder1]{\pic{encoder={\tiny q}{pooler}{pooler_color}{pooler_outline_color}};\\};

\node[matrix, inner sep=0] (passage_encoder3) [right =3.0 mm of query_encoder1]{\pic{encoder={\tiny d}{pooler}{pooler_color}{pooler_outline_color}};\\};
\node[matrix, inner sep=0] (query_encoder3) [right =1.0 mm of passage_encoder3] {\pic{encoder={\tiny q}{pooler}{pooler_color}{pooler_outline_color}};\\};

\node[matrix, outer sep=0, inner sep = 0] [right = 3.0mm of query_encoder3] (cross_encoder) {\pic {crossEncoder};\\};
\node[matrix, outer sep=0, inner sep = 0] (score4) [above = 2.0mm of cross_encoder] {\pic {score};\\};
\draw[->, line width=0.2mm, color=gray] (cross_encoder.north) -- (score4.south);

\node[matrix, inner sep=0] (passage_encoder2) [right =3.0 mm of cross_encoder]{\pic{encoder={\tiny d}{pooler}{pooler_color}{pooler_outline_color}};\\};
\node[matrix, inner sep=0] (query_encoder2) [right =1.0 mm of passage_encoder2] {\pic{encoder={\tiny q}{\HyperHeadName{}}{hyper_head_color}{hyper_head_outline_color}};\\};

\node[rectangle, outer sep=0, inner sep=1, draw=gray, rounded corners=0.5mm, minimum size=3mm] 
(ip_block) at ($(query_encoder1)!0.5!(passage_encoder1) + (0,1.0)$) {IP};
\node[matrix, outer sep=0, inner sep = 0] (score1) [above = 2.0mm of ip_block] {\pic {score};\\};

\draw[->, line width=0.2mm] (passage_encoder1.north) to[out=90,in=180] node[left, align=center, font=\tiny, rotate=0] {$E_d$} (ip_block.west);
\draw[->, line width=0.2mm] (query_encoder1.north) to[out=90,in=0] node[right, align=center, font=\tiny, rotate=0] {$E_q$} (ip_block.east);
\draw[->, line width=0.2mm, color=gray] (ip_block.north) to (score1.south);

\node[matrix, outer sep=0, inner sep=1, draw=gray, rounded corners=0.5mm] (q_net) at ($(query_encoder2)!0.5!(passage_encoder2) + (0,1.2)$) {\pic {qnet};\\};
\node[matrix, outer sep=0, inner sep = 0] (score2) [right = 2.0mm of q_net] {\pic {score};\\};

\draw[->, line width=0.2mm] (passage_encoder2.north) to[out=90, in=180] node[left, align=center, font=\tiny, rotate=0] {$E_d$} (q_net.west);
\draw[->, line width=0.2mm, color=gray] (query_encoder2.north) to[out=90, in=270]  (q_net.south);
\draw[->, line width=0.2mm, color=gray] (q_net.east) to (score2.west);

\node[font=\tiny, yshift=2mm, align=center] [left = 0.0mm of q_net] {input};
\node[font=\tiny, yshift=-0.8mm, align=center] [above = 0.0mm of q_net] {q-net};

\node[rectangle, outer sep=0, inner sep=1, draw=gray, rounded corners=0.5mm, minimum height=2mm, font=\tiny] 
(combine_block) at ($(query_encoder3)!0.5!(passage_encoder3) + (0,0.8)$) {combine};
\draw[->, line width=0.2mm] (passage_encoder3.north) to[out=90, in=180] node[left, align=center, font=\tiny, rotate=0] {$E_d$} (combine_block.west);
\draw[->, line width=0.2mm,] (query_encoder3.north) to[out=90, in=0] node[right, align=center, font=\tiny, rotate=0] {$E_q$}  (combine_block.east);

\node[rectangle, outer sep=0, inner sep=1, draw=gray, rounded corners=0.5mm, minimum height=2mm, font=\tiny] 
(learned_sim) at ($(query_encoder3)!0.5!(passage_encoder3) + (0,1.12)$) {MLP};

\draw[->, line width=0.2mm,] (combine_block.north) to[out=90, in=-90] (learned_sim.south);

\node[matrix, outer sep=0, inner sep = 0] (score3) [above = 1.4mm of learned_sim] {\pic {score};\\};
\draw[->, line width=0.2mm, color=gray] (learned_sim.north) -- (score3.south);

\node[align=center, font=\tiny] at ($(query_encoder1)!0.5!(passage_encoder1) + (0.0,-0.7)$) {Vector Similarity};
\node[align=center, font=\tiny \bf] at ($(query_encoder2)!0.5!(passage_encoder2) + (0.0,-0.7)$) {\name{}};
\node[align=center, font=\tiny] at ($(query_encoder3)!0.5!(passage_encoder3) + (0.0,-0.7)$) {Learned Similarity};
\node[align=center, font=\tiny ] at ($(cross_encoder)!0.5!(cross_encoder) + (0.0,-0.7)$) {Cross-encoding Similarity};

\end{tikzpicture}
}
\caption{Overview comparing \name{} to existing retrieval and reranking paradigms. \textcolor{gray}{Gray} arrows indicate the arrow does not represent an entity, it is the same as what it points to, in contrast black arrows do indicate a unique entity which is always labeled.}
\label{fig:main_overview}
\end{figure*}
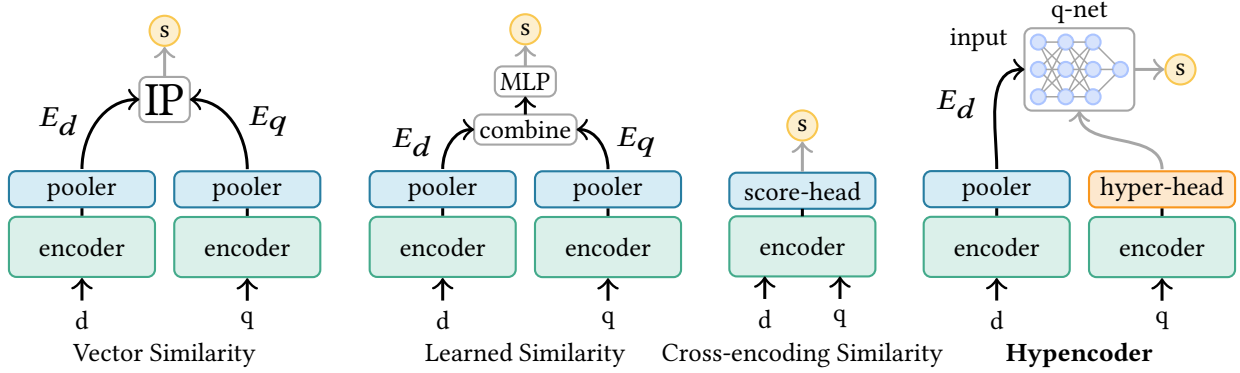

\section{Related Work}
\paragraph{\textbf{Vector Space Models}}
Vector-based models that use sparse vectors have existed for decades, with each index representing a term in the corpus vocabulary. Document-query similarity is computed using measures like $\ell^2$ distance, inner product, or cosine similarity, with term weighting methods such as TF-IDF being a substantial focus to improve performance. With the emergence of deep neural networks, focus shifted to learning representations for queries and documents. SNRM by \citet{SNRM} was the first deep learning model to retrieve documents from large corpora by learning latent sparse vectors. Following works leveraged pretrained transformer models like BERT \cite{BERT} using single dense vector representations \cite{DPR}. Recent improvements have focused on training techniques including self-negative mining \cite{ANCE, RocketQA, RANCE, adore, uia}, data augmentation \cite{RocketQA, DRAGON}, distillation \cite{MarginMSE, DRAGON, TCT-ColBERT, lion}, corpus-pretraining \cite{CoCondensor, Contriever, RetroMAE,b-prop}, negative-batch construction \cite{TAS-B} and curriculum learning \cite{CL-DRD,prod}. Alternative approaches include ColBERT \cite{ColBERT_v1}, which uses multiple dense vectors, and SPLADE \cite{SPLADE}, which revisits sparse representations using pretrained masked language models.

Though these methods vary substantially, they all share a fundamental commonality, that relevance is based on an inner product (or in some cases cosine similarity). We believe that this is a significant limitation of these methods and one which hampers the performance of these models on complex retrieval tasks. Our method circumvents this limitation by learning a small query-dependent  neural network that is fast enough to run on the entire collection (or approximately run; see Section \ref{sec:EfficientSearch} for details).

\paragraph{\textbf{Learned Relevance Models}}

Lightweight relevance models using neural networks have demonstrated improved retrieval performance compared to simple methods like inner products. Early iterations came in the form of learning-to-rank models \cite{RankNet, LearningToRankNonSmooth} which use query and document features to produce relevance scores for reranking. While these models traditionally used engineered features, more recent approaches adopted richer inputs. For instance, MatchPyramid \cite{MatchPyramid} and KNRM \cite{KNRM} use similarity matrices between non-contextualized word embeddings, while Duet \cite{Duetv1, DuetV2} combines sparse and dense term features in a multi-layer perceptron. DRMM \cite{DRMM} utilized histogram features as input to neural networks for scoring. Since the advent of BERT \cite{BERT}, focus has shifted to making transformer models more efficient, such as PreTTR \cite{PreTTR} which separately precomputes query and document hidden states. Recently, LITE \cite{LITE} extended ColBERT's similarity using column-wise and row-wise linear layers for scoring.

In the recommender systems community, learned similarity measures have been widely used \cite{LearnedCollaborativeFiltering, NeuralFactorizationMachines}. The common usage of neural scoring methods in recommendation has inspired research into efficient retrieval with more learned scoring signals. For instance, BFSG \cite{NeuralNetworkFastItemRanking} supports efficient retrieval with arbitrary relevance functions by using a graph of item representations and a greedy search strategy. A recent improvement on BFSG uses the scoring models gradient to prune directions that are unlikely to have relevant items \cite{GradientPruningTowardFastNeuralRanking}. Other works make use of queries to form a query-item graph to produce more informative neighbors \cite{FastNeuralRankingOnBipartiteGraphIndices}.

Our work differs from these works in one major way, the weights of neural scoring model are dynamically produced for each query. This means the full capacity of the scoring model can be dedicated to just the features which are meaningful for that specific query and that no additional combination step is needed to aggregate the query and document representations as the query representation is already encoded in the scoring model. Furthermore, the flexibility of our framework means we can replicate any existing learned relevance model as discussed in Section~\ref{sec:ComparisonExistingNeuralScoring}. On a broader note, there has been surprisingly little work on neural-based scoring for full-scale retrieval, especially in the transformer era. We hope our work can be a useful foundation and proof-of-concept for future work in this area.

\paragraph{\textbf{Hypernetworks}}
Hypernetworks also known as hypernets are neural networks which produce the weights for other neural networks. The term was first used by \citet{HyperNetworks} who demonstrated the effectiveness of hypernetworks to generate weights for LSTM networks. Since then, hypernetworks have been used in a variety of ways including neural architecture search \cite{GraphHyperNetworksNeuralArchitectureSearch}, continual learning \cite{ContinualLearningWithHypernetworks}, and few-shot learning \cite{MetaLearningWithLatentEmbeddingOptimization, HyperShot} to name a few. Generally, hypernetworks take a set of input embeddings that provide information about the type of task or network where the weights will be used. These embeddings are then projected to the significantly larger dimension of the weights of the ``main'' network. As the outputs of most hypernetworks are so large the hypernetworks themselves are often very simple such as a few feed-forward layers in order to keep computation feasible. Our case is unique in that our hypernetwork, the \name{}, is much larger than the small scoring network which we call \mininame{} (i.e. the ``main'' network). Additionally, to the best of our knowledge, this paper represents the first work to explore hypernetworks for first stage retrieval.

\section{\name}

Neural ranking models can be generally categorized into early-interaction and late-interaction models \cite{Dehghani2017WeakSupervision, MonoBERT, ColBERT_v1, SimLM,citadel,coil,colberter}. Currently, the most common implementation of early-interaction models is in the form of \textit{cross-encoders} (Figure~\ref{fig:main_overview} (second from the left)), where the query text $q$ and document text $d$ are concatenated (together with some predefined tokens or templates) and fed to a transformer network that learns a joint representation of query and document and finally produces a relevance score. The joint representation prevents these models from being able to precompute document representations, so they cannot be used efficiently on large corpora \cite{Guo2020,Mitra:2018:NeuralIR}. 

The most popular implementation of late-interaction models follows a \textit{bi-encoder} (or two tower) network architecture (Figure~\ref{fig:main_overview} (left)), where query and document representations are computed separately and a scoring function is used to estimate the relevance score. Formally, let $E_q \in \mathbb{R}^{n \times h}$ denote the representation learned for query $q$ consisting of $n$ $h$-dimensional vectors. Similarly, $E_d \in \mathbb{R}^{m \times h}$ denotes the representation learned for document $d$ consisting of $m$ vectors of the same dimensionality. The relevance score between $q$ and $d$ is computed as follows:
\begin{equation}
    \psi(E_q, E_d)
\end{equation}
where $\psi: \mathbb{R}^{n \times h} \times \mathbb{R}^{m \times h} \rightarrow \mathbb{R}$ denotes the scoring function. 

In order to take advantage of efficient indexing techniques, such as an inverted index or approximate nearest neighbor (ANN) search, many existing works use pooling techniques to obtain a single vector representation for each query and document and then employ simple and lightweight scoring functions, such as inner product or cosine similarity. There also exist more expensive methods that do not use pooling and perform such lightweight scoring functions at the vector level and then aggregate them, such as the maximum inner product similarity used in ColBERT \cite{ColBERT_v1}.

\paragraph{\textbf{On the Limitations of Linear Similarity Functions such as Inner Product}}
We believe the simple similarity functions used by existing bi-encoder models are not sufficient for modeling complex relationships between queries and documents. These functions inherently limit retrieval models to judge relevance in a way that can be represented by an inner product. Furthermore, it has been shown that the ability to compress and reconstruct information is correlated with the size, and thus complexity, of neural models \cite{LanguageModelingIsCompression}. This result indicates that using a relevance function as simple as an inner product likely reduces the amount of information that can be stored in a fixed representation size.
These factors explain why state-of-the-art dense retrieval models continue to underperform cross-encoder models, in terms of retrieval quality \cite{lin2022pretrained}. In the following, we show the limitations of inner products (as a linear similarity function) by theoretically demonstrating the impossibility of inner products to produce perfect rankings for some queries, regardless of the method used to create the query and document embeddings.

Let $C$ denote a corpus of $N$ documents, each represented by an $h$-dimensional vector. A perfect ranking of documents in $C$ for a provided query is a ranking where all relevant documents are ranked above all non-relevant documents. According to Radon's Theorem \cite{RadonTheorem}, any set of $h+2$ document vectors of dimension $h$ can be partitioned into two sets whose convex hulls intersect. An important application of Radon's Theorem is in calculating the Vapnik–Chervonenkis (VC) dimension \cite{VCDim} of $h$-dimensional vectors with respect to linear separability. For any $h+2$ vectors, the two subsets of a Radon partition cannot be linearly separated. In other words, for $N>h+1$, there exists at least one group of documents that is not linearly separable from the rest. In the real world, since $N \gg h+1$, there are indeed many such non-separable subsets. If any two of these subsets contain all the relevant documents for a query, then no linear similarity function can perfectly separate relevant from irrelevant documents. This includes inner product similarity and guarantees, that for some query, there will be an imperfect ranking.

To overcome these limitations with inner product similarity we use a multi-layer neural network with query-conditioned weights as our similarity measure. As neural networks are universal approximators \cite{UniversalApproximators}, \name{}'s similarity function can express far more complex functions than those expressed by inner products. A related alternative approach with the same benefits takes the query and document representations, combines them (e.g., through concatenation or similarity matrices), and feeds them to a neural network to serve as a similarity function (Figure~\ref{fig:main_overview} (second from the right)). However, this approach suffers from the following shortcomings: (1) query and document representations now need to be combined before scoring -- adding latency proportional to the complexity of the method used to combine them; (2) having separate query and document representations increases the input dimension to the neural network further increasing latency; (3) for efficiency reasons, the query representation is often pooled or compressed before being input into network which reduces the information the model receives. \name{} addresses these shortcomings. Since the query is directly encoded as the neural network's weights no concatenation or other form of combining inputs is needed, the document representation can be directly input to the scoring network. This, in addition to the reduced network size from having only document representations as input, allows for a substantial latency improvement. Further, as \name{} produces a query-specific neural network, every weight can be used to store query-related information without any need for compression or additional overhead. Lastly, we show in Section~\ref{sec:ComparisonExistingNeuralScoring} that existing learned relevance methods can be exactly replicated by \name{} with the additional flexibility of learning query-specific weights when desirable.

\subsection{\name{} Overview}

An overview of our model is depicted in Figure~\ref{fig:main_overview} (right); it represents a new category of models that sit between a cross-encoder and a bi-encoder model. Like a bi-encoder model, our method computes the query and document representations separately, but unlike most existing retrieval methods, our method allows for more complicated matching signals like those present in cross-encoder models. Following existing methods, we have a query encoder and a document encoder. When a document $d$ is input into the document encoder, we obtain a representation similar to existing encoder models, namely a set of one or more vectors $E_d \in \mathbb{R}^{m \times h}$ that represent the document's content, where $m$ is the number of vectors and $h$ is the dimension of the vectors. Although we focus on vectors in this work, in theory, the representation can be anything a neural network can output.

Now comes our unique contribution, which allows our method to consider more complex similarity signals. Given the query $q$, the query encoder \QueryEncoderName{} first produces a set of contextualized embeddings in a similar way to existing encoder models which we will call $E_q \in \mathbb{R}^{n \times h}$, where $n$ is the number of embeddings and $h$ is the dimension of the embeddings. At this point, existing methods apply a simple pooling mechanism; in contrast, our query encoder uses a \HyperHeadName{}. \textit{The \HyperHeadName{} takes $E_q$ and produces a set of matrices and vectors that are then used as the weights and biases for a small neural network which we coin the \mininame{}.} The \mininame{} is a query-dependent function to estimate relevance scores for a document, meaning each \mininame{} is unique to the query that created it, unlike existing neural scoring methods, which use a shared set of weights for all queries. To find the relevance of a document, the document representation $E_d$ is passed as input to the \mininame{} which outputs the relevance score.

\name{} is a generic framework that allows direct application of existing paradigms from neural retrieval and, more broadly, machine learning. For example, \name{} could easily work with multiple vectors similar to existing multi-vector models \cite{ColBERT_v1}, or use training routines popularized in dense retrieval, e.g., \cite{ANCE,RANCE,CL-DRD,DRAGON}. As an initial exploration, this paper focuses on showing the efficacy of \name{} without additional complexity, and thus uses a single vector document representation and no complex training recipes.

\subsection{Query and Document Encoders}
The \name{} framework is generic and can be applied to any implementation of query and document encoders. In this work, we use pretrained transformer-based encoder models commonly used in the recent neural network literature. Specifically, we use a pretrained BERT base model \cite{BERT} for encoding queries and documents. Although \name{} can operate on all document token representations this work focuses on a single vector representation of documents, which is more efficient in terms of query latency, memory requirements, and disk usage. To do so, we can either use the contextualized embedding representing the \texttt{[CLS]} token or take the mean of all the contextualized embeddings for all the non-pad input tokens. Empirically, we find that using the \texttt{[CLS]} token performs better. Therefore, the document representation produced by the encoder is a single vector with 768 dimensions, i.e., the same as BERT's output dimensionality. We refer to it as $E_d \in \mathbb{R}^{m \times h}$, where $m=1$ in our setting. 

Since \name{} only uses the contextualized-query-token representations once to produce the \mininame{}, it can skip pooling tokens without adding much cost. Therefore, we use all non-pad-token representations produced by the query encoder as the intermediate representation of the queries, denoted by $E_q \in \mathbb{R}^{n \times h}$, where $n$ is the number of tokens in the query $q$ and $h$ is the embedding dimensionality ($h=768$ in BERT).

\subsection{The Hyperhead Layers} \label{sec:hyperhead-layers}
The method to transform $E_q$ into the weights and biases for the \mininame{} is performed by the \hyperheadlayer{}s and is completely flexible. During our experimentation, we tried two mechanisms to do this transformation as well as many minor variants and found them all to have stable training, which suggests the \name{} framework is robust to the exact \hyperheadlayer{} implementation. Though we tried two approaches, we settled on one for the final set of experiments in this paper which we will now describe.

In the finalized implementation, the contextualized query embeddings $E_q \in \mathbb{R}^{n \times h}$, produced by the query encoder, are independently transformed by $l$ \hyperheadlayer{}s where $l$ is the number of weight matrices and bias vectors needed by the \mininame{}. For improved clarity, we focus only on the weight creation process for
$W_i^q \in \mathbb{R}^{r \times t}$ which is the weight matrix for the $i$\textsuperscript{th} layer of the \mininame{}. The values of $r$ and $t$ are selected based on the desired shape of the final weight matrix. The first step is to turn $E_q$ into key and value matrices:
\begin{equation}
    K^q_i = [E_q; 1] \times \theta_{K_i} \quad\quad V^q_i = [E_q; 1] \times \theta_{V_i} 
\end{equation}
where $\theta_{K_i}, \theta_{V_i} \in \mathbb{R}^{(h + 1) \times t}$ denote learnable parameters for constructing key and value matrices. In the above equation, the embedding matrix $E_q$ is concatenated with a column of all ones (i.e., $[E_q; 1]$) to model both weight multiplication and bias addition.

With the keys and values in hand, single-head scaled-dot-product attention \cite{AttentionIsAllYouNeed} is performed using a query matrix $Q_i \in \mathbb{R}^{r \times t}$. The matrix $Q_i$ is a set of learnable embeddings, similar to those used as input tokens for transformer models. Hence, the hidden layer representation $H_i \in \mathbb{R}^{r \times t}$ is computed as follows:
\begin{equation}
    H_i = \text{softmax} \left(\frac{Q_i K_i^T}{\sqrt{h}} \right) V_i
\end{equation}

A ReLU activation \cite{ReLU} is then applied to each $H_i$ followed by layer normalization \cite{LayerNorm} along the last dimension. Next a point-wise feed-forward layer is applied to produce $\widehat{H}^q_i \in \mathbb{R}^{r \times t}$:
\begin{equation}
    \widehat{H}^q_i = \theta_{W_i} \text{L-Norm}\left(\text{ReLU}(H_i)\right) + \theta_{b_i}
\end{equation}
where $\text{L-Norm}$ denotes layer normalization and $\theta_{W_i} \in \mathbb{R}^{t \times t}$ and $\theta_{b_i} \in \mathbb{R}^{t \times 1}$ are learnable parameters. Note, there are no learnable parameters in the layer normalization.

The final operation to get the $i$\textsuperscript{th} weight $W^q_i$ for \mininame{} is:
\begin{equation}
    W^q_i = \widehat{H}^q_i + \theta_{H_i}
\end{equation}
where $\theta_{H_i} \in \mathbb{R}^{r \times t}$ is the same size as $\widehat{H}^q_i$ and acts as a base weight which allows the model to learn universal (i.e., query-independent) patterns that are applicable for all queries. 

The process for the bias vectors is identical except the query matrix used in the attention operation $Q_i \in \mathbb{R} ^ {r \times t}$ has $r=1$ as there is only a single column in the output. 

\subsection{The \mininame{} Network}

The weights and biases produced by the \hyperheadlayer{}s are not by themselves a neural network. They need a certain arrangement and additional components (e.g. non-linearity). This is where the \name{}'s \mininame{} converter comes in. The converter knows the architecture of the \mininame{} and given the weights and biases from the \hyperheadlayer{}s, it produces a callable neural network object which takes as input the document representation $E_d$. 

It is worth highlighting that because the \mininame{}'s architecture is not strictly tied to how the \hyperheadlayer{} produces the weights and biases, it is simple to modify the architecture of the \mininame{}. All the \hyperheadlayer{}s need to know is how many weights and biases are needed and what shape they should be.

We use a simple feed-forward architecture for the \mininame{}. The output $x^d_{i+1}$ for the input $x^d_i$ at a given layer $i$ is given by:
\begin{equation}
    x^d_{i+1} = \text{L-Norm}\left(\text{ReLU}(W^q_i(x^d_i) + b^q_i)\right) + x^d_i,
\label{eq:mininet-layer}
\end{equation}
where $\text{L-Norm}$ represents a layer normalization without learnable parameters and the addition of $x^d_i$ is a residual connection. No residual connection is applied before the final layer. The layer in Equation~\eqref{eq:mininet-layer} is repeated until the last layer where a relevance score is produced using a linear projection layer with an output dimensionality of 1.

\subsection{Training}
Training \name{} is no different from training a bi-encoder as it shares the same core components, i.e., a query encoder and document encoder. The only difference is that instead of using an inner product to find the similarity, the \mininame{} is applied to the document representations. Thus, our novel contributions are focused solely on the architecture and not on a specific training technique. In this paper we employ a simple distillation training setup, for more details see Section~\ref{sec:ExperimentalSetup}.

\subsection{Efficient Retrieval using \name{}} \label{sec:EfficientSearch}
Being able to perform efficient retrieval is crucial for many real-world search scenarios where an exhaustive search is not feasible. For \name{} models, there is a clear parallel to dense models as both represent documents as dense vectors, but the differences between \name{} and dense models make it unclear whether the same efficient search techniques will work. For instance, it is clear that due to the linear nature of inner products, similar document vectors are likely to have similar inner products with a query vector; in the case of \name{} this assumption may not hold true as the non-linear nature of the \name{} scoring function could mean small differences in the input vector produce significant differences in the output score. 

To study the extent to which \name{}'s retrieval can be approximated for efficient retrieval, we developed an approximate search technique based loosely on navigating small world graphs \cite{SmallWorld, HNSW}. In the indexing stage we construct a graph where documents are nodes connected to their neighbors by edges.
We use $\ell^2$ distance between document embeddings similar to \citeauthor{NeuralNetworkFastItemRanking} \cite{NeuralNetworkFastItemRanking}.

After constructing the document graph, approximate search is performed following Algorithm~\ref{alg:efficient_search}. In brief, a set of initial candidate documents $\Tilde{C}$ is selected at random, these candidates are scored with the \mininame{} (line~\ref{alg:line:find_top}) and in lines \ref{alg:line:16}-\ref{alg:line:19} the best $nCandidates$ and their neighbors become the next candidates. In lines \ref{alg:line:12}-\ref{alg:line:15}, the top scoring candidates are added to $T$--a set which stores the $k$ best scoring documents so far. The algorithm terminates when one of three conditions is met: (1) the number of iterations equals $maxIter$; see line \ref{alg:line:4}, (2) there are no more candidates; see line \ref{alg:line:4}, or (3) no new documents are added to $T$ at a given step; see line \ref{alg:line:8}. We also consider an option without the final termination condition which we call \textit{without early stopping}. As the number of operations is dependent on the number of initial candidates $|\Tilde{C}|$, the running time is not tied to the number of documents, resulting in a run time complexity of $O(|\Tilde{C}| + nCandidates \cdot  maxIter)$. 

With this algorithm, we found that \name{} is able to significantly increase retrieval speed without a large loss in quality. See the results in Section~\ref{sec:AnalysisOfEfficiency}.

\begin{algorithm}[t]
\caption{\name{} Efficient Search.}%
\label{alg:efficient_search}

\begin{algorithmic}[1]
\Statex \textbf{Input:} \mininame{} $q$, \#NN to return $k$, initial candidates $\Tilde{C}$, \# candidates to explore every iteration $nCandidates$, $maxIter$
\Statex \textbf{Output:} $k$ closest neighbors to $q$
\State $v \leftarrow \Tilde{C}$ \Comment{set of visited elements}
\State $T \leftarrow \{-\infty\}$ \Comment{Stores top $k$ nearest neighbors to $q$ at any given time}
\State $i \leftarrow 0$ \Comment{Current iteration}
\While{$|\Tilde{C}| > 0$ and $i < maxIter$} \label{alg:line:4}
    \State $c \leftarrow$ find top $nCandidates$ values in $\Tilde{C}$ using $q$ \label{alg:line:find_top}
    \State $f \leftarrow$ get lowest scoring element from $T$
    \If{$\max_{\hat{c} \in c} \hat{c} < f$}
        \State \textbf{break} \Comment{all candidates are worse than $T$ so stop now} \label{alg:line:8}
    \EndIf
    \State $\Tilde{C} \leftarrow \{\}$ \Comment{Reset $\Tilde{C}$}
    \For{each $e \in c$} \label{alg:line:10}
        \State $f \leftarrow$ get lowest scoring element from $T$ 
        \If{$q(e) > f$ or $|T| < k$} \label{alg:line:12}
            \State $T \leftarrow T \cup e$
            \If{$|T| > k$}
                \State $T \leftarrow T \setminus \{f\}$   \label{alg:line:15}
            \EndIf
        \EndIf

        \For{each $n \in $ NEIGHBORS($e$)} \label{alg:line:16}
            \If{$n \notin v$}
                \State $\Tilde{C} \leftarrow \Tilde{C} \cup n$
                \State $v \leftarrow v \cup n$ \label{alg:line:19}
            \EndIf
        \EndFor
    \EndFor
    \State $i \leftarrow i + 1$
\EndWhile
\State \Return $T$
\end{algorithmic}
\end{algorithm}

\subsection{Comparison to Existing Neural IR Methods} \label{sec:ComparisonExistingNeuralScoring}

We argue that \name{} can exactly reproduce existing neural ranking models.
Let us start by formalizing the main components of existing neural methods: (1) a query representation $E_q$, (2) a document representation $E_d$, (3) some combination function $f_c(\cdot, \cdot)$, and (4) the final neural network that produces a score $f_s(\cdot)$. In comparison, \name{} 
does not have an $f_c(\cdot, \cdot)$ as the \mininame{} takes $E_d$ as its only input. 
We will now demonstrate that \name{} can exactly replicate any neural retrieval method that has the components above. The first step is to include $E_q$ and $f_c(\cdot, \cdot)$ in the \mininame{}. This allows the \mininame{} to exactly produce the input to $f_s(\cdot)$ that the existing neural retriever used. Next we reproduce the neural function from $f_s(\cdot)$ with query-dependent weights in the \mininame{}. When $E_d$ is input to the \mininame{}, all the original components of the existing neural retrieval model are present and thus the score can be exactly replicated. There is one difference, which is the weights of $f_s(\cdot)$ are query-dependent. However, this can be remedied in two ways: (1) shared weights can be used for all queries exactly replicating the original neural method (2) the weights for $f_c$ can have a common query-independent base weight, similar to our implementation (see details in Section~\ref{sec:hyperhead-layers}), this way if query-dependent weights are not beneficial the shared weight can be used, but if there are additional benefits the model and optimizer can learn to take advantage of them. Thus, \name{} can not only exactly replicate all existing neural retrieval methods it also allows the model to dynamically leverage query-dependent weights when the model determines they are beneficial.

\section{Experiments}
\subsection{Datasets}
\subsubsection{Training Dataset} \label{training_datasets}
For training our models we used the MSMARCO passage retrieval dataset \cite{MSMARCO} which contains 8.8M passages and has 503K training queries with at least one relevant passage. The queries in the MSMARCO training set are short natural language questions asked by users of the Bing search engine.

To create the training pairs, we first retrieved 800 passages for every query using an early iteration of \name{}. From these, we sampled 200 passages—the top 100 passages and another 100 randomly sampled from the remaining 700 passages. These query-passage pairs were then labeled using the MiniLM cross-encoder\footnote{Available at \url{https://huggingface.co/cross-encoder/ms-marco-MiniLM-L-12-v2}.} from the Sentence Transformers Library \cite{SentenceBERT}. %

\subsubsection{Validation Dataset}
For validating and parameter tuning, we use the TREC Deep Learning (DL) 2021 \cite{TREC-DL-21} and 2022 passage task \cite{TREC-DL-21, TREC-DL-22}. As the passage collection for TREC DL '21 and '22 is large and we wanted validation to be fast we created a subset with only passages in the QREL files. 

\subsubsection{Evaluation Datasets}
Our evaluation explores retrieval performance in three different areas: in-domain performance, out-of-domain performance, and performance on hard retrieval tasks. 

For in-domain performance, we use the MSMARCO Dev set \cite{MSMARCO}, TREC Deep Learning 2019 \cite{TREC_DL_2019}, and TREC Deep Learning 2020 \cite{TREC_DL_2020}. The MSMARCO Dev set contains around 7k queries with shallow labels, the majority of queries only have a single passage labeled as relevant. This collection uses queries from the same distribution as the training queries making it a clear test of the in-domain performance. On this dataset we report the standard evaluation metrics: MRR and Recall@1000. The TREC Deep Learning 2019 and 2020 datasets have a similar query distribution to MSMARCO Dev but feature far fewer queries (97 queries combined). The lower number of queries is compensated by far deeper annotations with every query having several annotated passages.

\pgfplotsset{ every non boxed x axis/.append style={x axis line style=-},
     every non boxed y axis/.append style={y axis line style=-}}

\definecolor{graph_color_1}{HTML}{277da1}
\definecolor{graph_color_2}{HTML}{f8961e}

\begin{figure*}[h!]
    \centering

    \begin{tikzpicture}[baseline]
    \begin{axis}[ 
        scale only axis, 
        axis y line*=left, 
        xlabel=$|\Tilde{C}|$, 
        ylabel=\textcolor{graph_color_1}{\footnotesize nDCG@10},
        width=0.19\textwidth,
        height=2cm,
        legend style={
            at={(1.0,-0.25)},
            anchor=north east, 
            text=black,
            legend image post style={black},
            nodes={scale=0.6},
        },
        enlarge y limits=0.1,
        ytick={0.64, 0.66, 0.68, 0.7, 0.72},
        xtick pos=left,
        ytick pos=left,
        ylabel near ticks,
        xlabel near ticks,
        xmode=log,
    ]
        \addplot[graph_color_1, line width = 1.0pt] coordinates {(10,0.657227173201666)(100,0.6839815941070432)(1000,0.6970931130909354)(10000,0.7028781904961429)(10000,0.7028781904961429)(10000,0.7028781904961429)(100000,0.6968752971431624)};
        \addplot[graph_color_1, dashed, line width = 1.0pt] coordinates {(10,0.657227173201666)(100,0.6839815941070432)(1000,0.7098415361176305)(10000,0.7156266135228379)(10000,0.7156266135228379)(10000,0.7156266135228379)(100000,0.6984160930374507)};
    \end{axis}
    \begin{axis}[
        scale only axis, 
        axis y line=right, 
        axis x line=none, 
        ylabel=\textcolor{graph_color_2}{\footnotesize Query Latency (ms)},
        width=0.19\textwidth,
        height=2cm,
        enlarge y limits=0.1,
        ylabel near ticks,
        xmode=log,
        xlabel near ticks,
    ]
        \addplot[graph_color_2, line width = 1.0pt] coordinates {(10,78.49962212318599)(100,65.96895151360091)(1000,63.07049684746321)(10000,57.6694178026776)(10000,57.6694178026776)(10000,57.6694178026776)(100000,143.09355824492698)};
        \addplot[graph_color_2, dashed, line width = 1.0pt] coordinates {(10,73.46584076105161)(100,80.34529796866484)(1000,81.01236542990041)(10000,90.0667434514955)(10000,90.0667434514955)(10000,90.0667434514955)(100000,176.44097084222838)};
    \end{axis} 
    \end{tikzpicture}
    \begin{tikzpicture}[baseline]
    \begin{axis}[ 
        scale only axis, 
        axis y line*=left, 
        xlabel=$nCandidates$, 
        ylabel=\textcolor{graph_color_1}{\footnotesize nDCG@10},
        width=0.19\textwidth,
        height=2cm,
        legend style={
            at={(1.0,-0.25)},
            anchor=north east, 
            text=black,
            legend image post style={black},
            nodes={scale=0.6},
        },
        enlarge y limits=0.1,
        ytick={0.64, 0.66, 0.68, 0.7, 0.72},
        xtick pos=left,
        ytick pos=left,
        ylabel near ticks,
        xlabel near ticks,
    ]
        \addplot[graph_color_1, line width = 1.0pt] coordinates {(8,0.6253591715706962)(16,0.6448268761626977)(32,0.6786175787253655)(64,0.7028781904961429)(64,0.7028781904961429)(64,0.7028781904961429)(128,0.6974597931411562)(256,0.7087882464950501)(512,0.7210713713340823)(1024,0.7341100378511684)};
        \addplot[graph_color_1, dashed, line width = 1.0pt] coordinates {(8,0.6253591715706962)(16,0.6448009750670014)(32,0.6785916776296691)(64,0.7156266135228379)(64,0.7156266135228379)(64,0.7156266135228379)(128,0.7137632491351856)(256,0.7215366695217451)(512,0.7341100378511684)(1024,0.7341100378511684)};
    \end{axis}
    \begin{axis}[
        scale only axis, 
        axis y line=right, 
        axis x line=none, 
        ylabel=\textcolor{graph_color_2}{\footnotesize Query Latency (ms)},
        width=0.19\textwidth,
        height=2cm,
        enlarge y limits=0.1,
        ylabel near ticks,
        xlabel near ticks,
    ]
        \addplot[graph_color_2, line width = 1.0pt] coordinates {(8,49.67966745066088)(16,50.005957137706666)(32,50.98761514175769)(64,57.6694178026776)(64,57.6694178026776)(64,57.6694178026776)(128,80.60029495594114)(256,141.22345835663552)(512,250.4059270370838)(1024,492.0011675635049)};
        \addplot[graph_color_2, dashed, line width = 1.0pt] coordinates {(8,61.17011225500772)(16,65.369306608688)(32,73.49416821501977)(64,90.0667434514955)(64,90.0667434514955)(64,90.0667434514955)(128,129.01445322258527)(256,236.37792675994163)(512,458.35791077724724)(1024,894.4306595380916)};
    \end{axis} 
    \end{tikzpicture}
    \begin{tikzpicture}[baseline]
    \begin{axis}[ 
        scale only axis, 
        axis y line*=left, 
        xlabel=$maxIter$, 
        ylabel=\textcolor{graph_color_1}{\footnotesize  nDCG@10},
        width=0.19\textwidth,
        height=2cm,
        legend style={
            at={(-0.35,-0.25)},
            anchor=north west, 
            text=black,
            legend image post style={black},
            nodes={scale=0.6},
        },
        enlarge y limits=0.1,
        ytick={0.3, 0.4, 0.5, 0.6, 0.7},
        xtick pos=left,
        ytick pos=left,
        ylabel near ticks,
        xlabel near ticks,
    ]
        \addplot[graph_color_1, line width = 1.0pt] coordinates {(2,0.26477399644530447)(4,0.6153305693793425)(8,0.6662349555809605)(12,0.7028781904961429)(12,0.7028781904961429)(12,0.7028781904961429)(16,0.7028781904961429)(20,0.7028781904961429)(24,0.7028781904961429)(28,0.7028781904961429)(32,0.7028781904961429)};
        \addplot[graph_color_1, dashed, line width = 1.0pt] coordinates {(2,0.26477399644530447)(4,0.6153305693793425)(8,0.6662349555809605)(12,0.7156266135228379)(12,0.7156266135228379)(12,0.7156266135228379)(16,0.7150811434696449)(20,0.715042117524844)(24,0.715042117524844)(28,0.715042117524844)(32,0.715042117524844)};
        \legend{w/ early stop, w/o early stop}
    \end{axis}
    \begin{axis}[
        scale only axis, 
        axis y line=right, 
        axis x line=none, 
        ylabel=\textcolor{graph_color_2}{\footnotesize Query Latency (ms)},
        width=0.19\textwidth,
        height=2cm,
        enlarge y limits=0.1,
        ylabel near ticks,
        xlabel near ticks,
    ]%
        \addplot[graph_color_2, line width = 1.0pt] coordinates {(2,28.885248095490212)(4,39.968363074369215)(8,55.21737143050793)(12,57.6694178026776)(12,57.6694178026776)(12,57.6694178026776)(16,57.981796042863714)(20,58.43507411868074)(24,61.809140582417335)(28,61.977292216101354)(32,59.12499095118323)};
        \addplot[graph_color_2, dashed, line width = 1.0pt] coordinates {(2,28.600265813428305)(4,39.628139761991285)(8,69.41637881966524)(12,90.0667434514955)(12,90.0667434514955)(12,90.0667434514955)(16,113.7952028318893)(20,137.69336633904035)(24,162.62342763501545)(28,187.94419044672057)(32,213.92392003258996)};
    \end{axis} 
    \end{tikzpicture}
    \caption{Relationship between the three main parameters of our efficient search: the size of the initial set of candidates $\Tilde{C}$, the number of neighbors to explore $nCanddidates$, and the number of iterations $maxIter$ and both effectiveness in terms of TREC DL '19 nDCG@10 and efficiency in terms of query latency in milliseconds.}
    \label{fig:approximate-search-tradeoff}
\end{figure*}
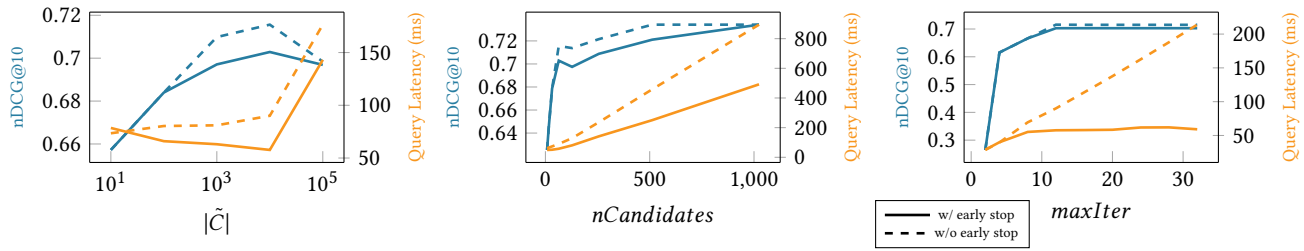

To assess out-of-domain performance, we evaluated question-answering tasks from various domains. Specifically, the TREC COVID \cite{TREC-Covid} and NFCorpus datasets \cite{NFCorpus} for the biomedical domain and FiQA \cite{FiQA} for the financial domain. We also evaluated on DBPedia \cite{DBpedia} as an entity retrieval dataset and on Touch\'{e} \cite{Touche} as an argument retrieval dataset. We use the BEIR \cite{BEIR} versions of these datasets from the ir\_datasets library.\footnote{Available at \url{https://ir-datasets.com/}.}

To explore the full capabilities of \name{} we want to evaluate how it performs on retrieval tasks that are more challenging than standard question-passage retrieval tasks. To some extent hardness is subjective, but we tried to define a clear set of criteria to define difficulty: (1) current neural retrieval models should struggle on the task, (2) term matching models like BM25 should also struggle on the task, (3) the queries are longer or otherwise more complicated than standard web queries. An additional requirement we had for tasks that were significantly different from the MSMARCO was training data to facilitate fine-tuning prior to evaluation. We believe this is reasonable as we are not investigating the models' zero-shot performance but the inherent limits of the model.

The first hard dataset is the TREC Tip-of-the-Tongue (TOT) Track from 2023 \cite{TREC-TOT-2023}. It contains queries written by users that know many aspects of the item they are looking for but not the name of the item. Thus, TOT queries tend to be verbose and can include many facets. The TREC TOT 2023 dataset specifically looks at TOT queries for movies with the corresponding movie's Wikipedia page as the golden passage. We used the 150 query development set as the test set relevance labels were not public at the time of writing. Each query has a single relevant passage. For training we use the data from  \citet{TOMT-TrainingDataset} which is around 15k TOT queries from Reddit for the book and movie domains. 

The second dataset is FollowIR \cite{FollowIR} for instruction following in retrieval. This dataset is built on top of three existing TREC datasets: TREC Robust '04 \cite{Robust04}, TREC News '21 \cite{TREC-News20}, and TREC Core '17 \cite{TREC-Core17}; it uses the fact that these datasets include instructions to the annotators which can act as a complex instruction. To test how well a retrieval system follows the instruction the creators of FollowIR modify the instruction to be more specific and re-annotate the known relevant documents. As training data we use MSMARCO with Instructions, a recent modification of MSMARCO which adds instructions to the queries as well as new positive passages and hard negative passages which consider the instruction \cite{MSMARCO-with-Instructions}.

The final dataset is a subset of TREC DL-HARD \cite{TREC_DL_HARD}. The full dataset uses some of the queries from TREC DL 2019 and 2020 as well as some queries that were considered for DL 2019 and 2020 but were not included in the final query collection. TREC DL-HARD is built specifically with the hardest queries from the TREC DL pool. The authors do so by using a commercial search engine to find queries that are not easily answered. The standard TREC DL-HARD dataset has 50 queries half of which appear in TREC DL 2019 or TREC DL 2020 and half of which are new queries which are labeled by the authors of TREC DL-HARD. We found that the queries labeled by the authors had far fewer judged documents in the top 10 documents compared to those labeled by TREC (around 15\% versus 93\%), this made the evaluation metrics unreliable so we decided to only use those with TREC labeling.

\subsection{Experimental Setup} \label{sec:ExperimentalSetup}
\subsubsection{Training Details}
All the \name{}s use BERT \cite{BERT} base uncased as the base model. We use PyTorch and Huggingface Transformers for model implementation and training. All of our \mininame{}s use an input dimension of 768 and a hidden dimension of 768. Unless otherwise stated, we use 6 linear layers in the \mininame{} not including the final output projection layer. 

We used a training batch size of 64 per device and 128 in total. A single example in the batch is a query, positive document, and 8 additional documents ranging in relevance. The positive document is the top ranked document by our teacher model. The other 8 documents are sampled randomly from the passages associated with the query. For more details about the dataset see Section \ref{training_datasets}. Passages were truncated to 196 tokens and queries to 32 tokens.

Our primary loss function is Margin MSE \cite{MarginMSE}. We construct (query, positive document, negative document) triplets where all of the negatives for a query form their own triplet. The loss is found by averaging the individual loss of all the triplets. In addition to Margin MSE, we use an in-batch cross entropy loss where the (query, positive document) is assumed to be a true positive and all the other queries' positive documents are assumed to be negatives. We do not consider the additional ``hard'' negatives from the query in the cross entropy loss as many of these documents are relevant to the query. We use AdamW as our optimizer with default settings and a learning rate of 2e-5 with a constant scheduler after a warm up of 6k steps. Our training hardware consists of two A100 GPUs with 80GB memory. Training took around 6 days.

To select the best model, we evaluate each model on the validation set every 50k steps and pick the model with the best R Precision within the first 800k steps. We selected R Precision because it balances both recall and precision in a single metric and does not require a predefined cutoff.

When training for the harder tasks we use AdamW with the learning rate 8e-6 with a linear scheduler and a warm-up ratio of 1e-1. We forgo Margin MSE and only use cross entropy loss. For TOT training we train for 25 epochs or 3.3k steps. For FollowIR training we train for 1 epoch or around 10k steps. We use a batch-size of 96. Each example in the batch includes a query, positive document, and hard negative document. We use a maximum document and query length of 512 tokens.

\subsection{Baselines}

For comparison with \name{}, we include several baseline models and models which we include for reference which are not directly comparable (see \ref{InDomainResults} for more details). Our main baselines which we evaluate on all datasets are TAS-B \cite{TAS-B}, CL-DRD \cite{CL-DRD}, BM25, and our own bi-encoder baseline which we call BE-Base. We train BE-Base exactly the same as \name{} except we use separate encoders and use a linear LR scheduler. We select our main dense baselines TAS-B and CL-DRD as they are both strong bi-encoder models which leverage knowledge-distillation training and which use the same document embedding dimension of 768. For in-domain results we include an additional set of dense models: ANCE \cite{ANCE}, TCT-ColBERT \cite{TCT-ColBERT}, and MarginMSE \cite{MarginMSE}. We only include these models in the in-domain results to save space in other sections and because TAS-B and CL-DRD outperform the other baselines. For reference we also include: the late-interaction model ColBERT v2 \cite{colbert_v2}; the neural sparse model SPLADE++ SD \cite{SPLADE++}; RepLLaMA a 7b parameter bi-encoder model \cite{RepLlama}; DRAGON a bi-encoder trained with 5 teacher models and 20M synthetic queries; MonoBERT a reranking model reranking the top 1k BM25 retrieavls \cite{MonoBERT}; and the reranking model cross-SimLM reranking the top 200 passages from bi-SimLM \cite{SimLM}. Reference results were taken from the RepLLaMA \cite{RepLlama} and DRAGON \cite{DRAGON} papers.

\subsection{Results and Discussion}

\begin{table}[t]
\caption{Comparison on in-domain evaluation datasets. The symbols next to each baseline indicate significance values with $p < 0.05$. Note that $\dagger$ is a group of baselines.} \label{table:in-domain-results}
\scalebox{0.82}{
\begin{tabular}{l!{\color{lightgray}\vrule}lll!{\color{lightgray}\vrule}ll!}
\toprule
                                    & \multicolumn{3}{c!{\color{lightgray}\vrule}}{\textbf{TREC-DL '19 \& '20}} & \multicolumn{2}{c}{\textbf{MSMARCO Dev}}  \\
Model                               & nDCG@10        & MRR             & R@1000         & MRR@10          & R@1000                    \\ 
\midrule
\multicolumn{6}{l}{\textbf{Single Vector Dense Retrieval Models \& BM25 (Baselines)}} \\
\textbf{BM25} $\dagger$  & 0.491          & 0.679          & 0.735          &               0.184 &      0.853                     \\
\textbf{ANCE} $\dagger$                 & 0.646          & 0.811          & 0.767          & 0.330          & 0.958                     \\
\textbf{TCT-ColBERT} $\dagger$ & 0.669          & 0.820          & 0.806          & 0.335          & 0.964                     \\
\textbf{Margin MSE} $\dagger$    & 0.669          & 0.845          & 0.782          & 0.325          & 0.955                     \\
\textbf{TAS-B} $\spadesuit$         & 0.700          & \uline{0.863}  & 0.861          & 0.344          & 0.978                     \\
\textbf{CL-DRD} $\diamondsuit$      & 0.701          & 0.844          & 0.838          & 0.382          & \textbf{0.981}            \\
\textbf{BE-Base} $\clubsuit$        & \uline{0.713}  & 0.855          & \uline{0.868}  & 0.359          & 0.980                     \\ 
\midrule
\textbf{\name{}} & \textbf{0.736}$^{\dagger \spadesuit \diamondsuit \clubsuit}$ & \textbf{0.885}$^{\dagger \diamondsuit}$ & \textbf{0.871}$^{\dagger \diamondsuit}$ & \textbf{0.386}$^{\dagger \clubsuit \spadesuit}$ & \textbf{0.981}$^{\dagger \spadesuit}$ \\
\midrule
\multicolumn{6}{l}{\textbf{Other Retrieval Models (Reference Models)}} \\
\textbf{ColBERT v2}                 & 0.749          & -              & -              & 0.397          & 0.984                     \\
\textbf{SPLADE++ SD}                & 0.723          & -              & -              & 0.368          & 0.979                     \\
\textbf{RepLLaMA}                   & 0.731          & -              & -              & 0.412          & 0.994                     \\
\textbf{DRAGON}                     & 0.734          & -              & -              & 0.393          & 0.985                     \\
\textbf{MonoBERT}                   & 0.722          & -              & -              & 0.372          & 0.853                     \\
\textbf{cross-SimLM}                & 0.735          & -              & -              & 0.437          & 0.987                     \\ 
\bottomrule
\end{tabular}
}
\vspace{-8pt}
\end{table}

\subsubsection{In-Domain Results} \label{InDomainResults}
Our in-domain results are presented in Table~\ref{table:in-domain-results}; they demonstrate that compared with baselines and even the reference models \name{} has very strong performance. \name{} is significantly better than each baseline in nDCG@10 on the combined TREC DL '19 and '20 and statistically better than all but CL-DRD on MSMARCO Dev RR@10. The most direct comparison, BE-Base, has far lower nDCG@10, RR, and RR@10 values indicating the \name{} is able to bring a large boost in precision based metrics over dense retrieval. In terms of recall \name{} is either as good or better than all the baselines though the gap is not as large as for precision based metrics. 

Impressively \name{} is able to surpass DRAGON on nDCG@10 on the combined TREC DL '19 and '20 query set, though DRAGON uses the same base model and is a bi-encoder, it uses 32 A100s to train, 40x the training queries, and a complex 5 teacher curriculum learning distillation training technique. In other words, DRAGON is likely close to if not the ceiling for BERT-based bi-encoders and still \name{} is able to match it with a simple distillation training setup and far less training compute.

\name{} also beats both rerankers MonoBERT and cross-SimLM; demonstrating that reranking cannot make up for a weak retriever's performance. Continuing in TREC-DL '19 and '20 we find that \name{} even surpassed RepLLaMA which is more than 60x larger and which also uses a significantly larger document embedding dimension of 4096. In fact the only model beating \name{} in nDCG@10 is ColBERTv2 which uses an embedding for every token in the document compared to \name{}'s fixed 768 dimension embedding. MSMARCO Dev results are also good with \name{} outperforming all the baselines and outperforming a few of the reference models such as SPLADE++ and MonoBERT.

Overall, \name{}'s in-domain results are exceptionally strong given the simple training routine used, small encoder model size, and minimal document representation size. To the best of our knowledge, \name{} sets a new record for combined TREC-DL '19 and '20 nDCG@10 with a 768 dimension dense document vector.

\subsubsection{Out-of-Domain Results}
Table~\ref{tab:out-of-domain} shows our results on the select out-of-domain datasets, we only include our main baseline models and BM25 due to space limitations. The general trend is that \name{} has strong out-of-domain performance in question answering tasks (Q\&A) and entity retrieval tasks. This indicates that despite \name{}'s more complex similarity function it is still able to generalize well in a zero-shot manner to new tasks.

\begin{table}
    \centering
    \caption{Out-of-domain results in nDCG@10. We only compare significance with BE-Base. Significance results with $p < 0.05$ are shown with $\BeBaseSymbol$ and $p < 0.1$ are shown with $\diamondsuit$.} \label{tab:out-of-domain}
    \scalebox{0.91}{
    \begin{tabular}{
        p{1.75cm}!{\color{lightgray}\vrule}llll!{\color{lightgray}\vrule}l!
    }
    \toprule
        & \multicolumn{4}{c!{\color{lightgray}\vrule}}{\textbf{Baselines}} & \textbf{Ours} \\
        \midrule 
        Rep type & sparse & dense & dense & dense & hypernet  \\
        
        \midrule
        & BM25 & TAS-B & CL-DRD & BE-Base & Hypecoder \\
        \midrule
        \multicolumn{2}{l}{\textbf{Q \& A }} \\
        TREC-Covid & \uline{0.656} & 0.481 & 0.584 & 0.651 & \textbf{0.688}$^{\diamondsuit}$ \\
        FiQA  & 0.236 & 0.300 & 0.308 & \uline{0.309} & \textbf{0.314} \\ 
        NFCorpus  & \textbf{0.325} & 0.319 & 0.315 & \uline{0.327} & 0.324 \\ 
        \midrule
        \multicolumn{2}{l}{\textbf{Misc.}} \\
        DBPedia  & 0.313 & 0.384 & 0.381 & \uline{0.405} & \textbf{0.419}$^{\BeBaseSymbol}$ \\
        Touché v2  & \textbf{0.367} & 0.162 & 0.203 & 0.240 & \uline{0.258}$^{\diamondsuit}$ \\
        \bottomrule
    \end{tabular}}
\vspace{-8pt}
\end{table}

\begin{table*}[h]
\caption{Evaluation metrics for the harder set of tasks which include TREC DL-HARD, TREC Tip-of-the-Tongue TOT, and FollowIR. Significance is shown at $p < 0.1$.} \label{tab:harder}
\scalebox{0.84}{
\begin{tabular}{l!{\color{lightgray}\vrule}ccc!{\color{lightgray}\vrule}ccc!{\color{lightgray}\vrule}cc!{\color{lightgray}\vrule}cc!{\color{lightgray}\vrule}cc}
\toprule
                 & \multicolumn{3}{c!{\color{lightgray}\vrule}}{\textbf{TREC DL-HARD}}       & \multicolumn{3}{c!{\color{lightgray}\vrule}}{\textbf{TREC TOT DEV}}       & \multicolumn{2}{l!{\color{lightgray}\vrule}}{\textbf{FollowIR Robust '04}} & \multicolumn{2}{l!{\color{lightgray}\vrule}}{\textbf{FollowIR News '21}} & \multicolumn{2}{l}{\textbf{FollowIR Core '17}}  \\
Model            & nDCG@10        & MRR             & R@1000         & nDCG@10        & MRR             & nDCG@1000      & AP             & p-MRR                   & nDCG@5         & p-MRR                 & AP             & p-MRR                 \\ 
\midrule
\textbf{BM25} $\BMSymbol{}$  & 0.466          & 0.813          & 0.646          & 0.086          & 0.088          & 0.131          & 0.121          & \textbf{-3.1}           & 0.193          & -2.1                  & 0.081          & \textbf{-1.1}                  \\
\textbf{TAS-B} $\TasBSymbol{}$   & 0.574          & 0.789          & 0.777          & 0.097          & 0.089          & 0.162          & 0.203          & -5.4                    & \uline{0.263}  & -0.8                  & 0.170          & -10.0                 \\
\textbf{CL-DRD} $\CLDRDSymbol{}$  & 0.573          & 0.790          & 0.719          & 0.088          & 0.082          & 0.151          & 0.206          & -7.2                    & 0.240          & \uline{-0.3}          & 0.162          & -12.1                 \\
\textbf{BE-Base} $\clubsuit$ & 0.607          & 0.864          & \textbf{0.805}          & \uline{0.121}  & \uline{0.110}  & \uline{0.179}  & \uline{0.207}  & -3.7                    & 0.239          & -1.1                  & \uline{0.178}  & \uline{-7.7}         \\ 
\midrule
\textbf{\name{}} & \textbf{0.630}$^{\BMSymbol{} \TasBSymbol{} \CLDRDSymbol{}}$ & \textbf{0.887}$^{\TasBSymbol{} \CLDRDSymbol{}}$ & \uline{0.798}$^{\BMSymbol{} \CLDRDSymbol{}}$ & \textbf{0.134}$^{\BMSymbol{} \TasBSymbol{} \CLDRDSymbol{}}$ & \textbf{0.125}$^{\TasBSymbol{} \CLDRDSymbol{}}$ & \textbf{0.182}$^{\CLDRDSymbol{} \BMSymbol{}}$ & \textbf{0.212}$^\TasBSymbol{}$ & -\uline{3.5}            & \textbf{0.272} & \textbf{2.0}        & \textbf{0.193} & -11.8       \\     \bottomrule
\end{tabular}
}
\vspace{-8pt}
\end{table*}

\subsubsection{Results on Harder Retrieval Tasks}
The results for the harder retrieval tasks are in Table~\ref{tab:harder}, like in the out-of-domain section we only consider the main baseline models and BM25. We can see that in the harder tasks \name{} remains dominant over the baseline models with higher retrieval metrics in the vast majority of metrics. Additionally, the relative improvement is higher compared to the in-domain results (on all metrics where \name{} is the best) suggesting that on harder tasks the added complexity that can be captured by \name{}'s similarity function is especially important. Additionally, the high performance on TREC tip-of-the-tongue (TOT) and FollowIR indicate that \name{} adapts well to different domains through domain-specific fine-tuning.

On the evaluated subset of TREC DL-HARD we see that \name{} has stronger precision metrics than the baselines by a large margin. As mentioned previously, the higher relative improvement suggests that \name{} is especially dominant on harder tasks which, in part, explains its higher performance on TREC DL '19 and '20. Though on in-domain dataset \name{} does better or the same on recall metrics, on TREC DL-HARD BE-Base has higher recall than \name{}. We suspect that this may be because the relevance function that the \mininame{} applies is not smooth, which has the benefit of being more discerning and likely accounts for some of the precision gains. However, if the \mininame{} makes a mistake the non-smooth scoring could result in a much harsher score than the linear inner product is capable of producing. 

Moving to TREC Tip-of-the-Tongue (TOT) we see that \name{} continues to perform well. Tip-of-the-tongue is a complex retrieval task with long multi-aspect queries and passages, the fact \name{} outperforms the baselines by a large margin validates the need for a more complex relevance function.

Finally, we have FollowIR which has three subsets -- on all three \name{} has the best performance on the retrieval evaluation metrics of choice, in many cases by a sizable margin. Beside the retrieval metrics, we also include p-MRR which is a metric introduced in the FollowIR \cite{FollowIR} paper. The metric measures the change in document ranks before and after an instruction is modified to see how well the model responds to the additional requirements. A p-MRR of 0 indicates no change based on the changed instruction and a p-MRR of +100 indicates the documents were perfectly changed while -100 indicates the opposite. As p-MRR is relative to each model's performance before the instructions are modified it is not indicative of stand-alone retrieval performance. That being said, \name{} is the only model to achieve a positive p-MRR indicating it correctly modified the document ranking based on the instruction. This is no small feat as the FollowIR paper reports no retrieval model with the same size of \name{} was able to get a positive p-MRR and even many much larger models trained on large instruction retrieval datasets could not get a positive result.

\subsubsection{Analysis of Efficiency} \label{sec:AnalysisOfEfficiency}
To be able to search large-scale collections \name{} has to work well while only doing computation on a small subset of the document corpus. This led us to develop the efficient algorithm in Section~\ref{sec:EfficientSearch}. To quantify the performance we do a set of experiments varying the key parameters to see how each impacts both search quality and query latency. The results of these experiments can be seen in Figure~\ref{fig:approximate-search-tradeoff}. Note that the document-to-document graph used has 100 neighbors per document. The figures show that each parameter has an important role in search quality. The largest role is played by $maxIter$ which can drastically reduce search performance if not set high enough, but which plateaus after around 12. The value of $nCandidates$ follows a similar pattern with a drastic increase followed by a plateau. The parameter $\Tilde{C}$, the number of initial candidates, is unique in that it has a more gradual rise and is the only parameter that causes a decrease in effectiveness if raised too high. We suspect this happens because only $nCandidates$ are explored at each iteration but all candidates in $\Tilde{C}$ are considered visited and thus are not explored in the future. This could mean that many good results from the initial candidates are not explored, leaving potentially good neighbors of these nodes undiscovered. Another interesting aspect of $\Tilde{C}$ is that time decreases as it increases when early stopping is used. This is likely because high quality candidates are found sooner allowing search to end more quickly. In general, our approximation seems to behave as expected in terms of both time and retrieval performance.

With the insights from our analysis above we developed two configurations, one which optimizes for speed and one that optimizes for retrieval quality. These two configurations compared against exhaustive search can be seen in Table~\ref{tab:ApproximateSearchResults}. Both approximate configurations significantly decrease the retrieval time when compared to the exhaustive approach.

\begin{table}
\centering
\caption{Average query latency and nDCG@10 on TREC DL '19 and '20 with efficient search. Efficient 1 uses parameters ($\Tilde{C}=10000$, $nCandidates=64$, $maxIter=16$), Efficient 2 uses parameters ($\Tilde{C}=100000$, $nCandidates=328$, $maxIter=20$). All model inference was performed on an NVIDIA L40S with BF16 precision.} 
\vspace{-0.3cm}\label{tab:ApproximateSearchResults}
\scalebox{0.85}{
\begin{tabular}{l!{{\color{lightgray}\vrule}}c!{\color{lightgray}\vrule}cc}
\toprule
\textbf{Search Type} & \textbf{Query Latency (ms)}  &\textbf{DL '19} & \textbf{DL '20}  \\
\midrule
Exhaustive & 1769.8 &  0.742 & 0.731                                    \\
Efficient 1 & 59.6  & 0.702 & 0.730                                  \\
Efficient 2 & 231.1 & 0.722  & 0.731                        \\
\bottomrule
\end{tabular}
}
\vspace{-8pt}
\end{table}

\subsubsection{Impact of \mininame{} Depth}
\name{} performance suggests that having a more complex relevance function does indeed help improve retrieval performance as we had hypothesized. This raises the question: what is the optimal complexity of this relevance function? To answer this question we trained four versions of \name{} each trained to produce a \mininame{} with a different number of layers. We selected \textit{[2, 4, 6, 8]} as the number of \mininame{} layers.

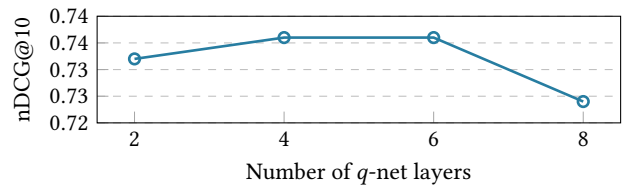
\begin{figure}
    \centering
    \begin{tikzpicture}
    \begin{axis}[
        title={},
        width=0.48\textwidth,
        height=3cm,
        xlabel={Number of \mininame{} layers},
        ylabel={nDCG@10},
        xmin=1.5, xmax=8.5,
        ymin=0.72, ymax=0.74,
        xtick={0,2,4,6,8},
        legend pos=south east,
        ymajorgrids=true,
        grid style=dashed,
        ylabel near ticks,
        xlabel near ticks,
        xtick pos=left,
        ytick pos=left,
    ]
    
    \addplot[color=graph_color_1, mark=o, line width=1.0pt]	
        coordinates {
        (2.0,0.732)(4,0.736)(6,0.736)(8,0.724)
        };
    
    \end{axis}
    \end{tikzpicture}
    \caption{Average nDCG@10 on TREC DL '19 and '20 versus number of layers in the \mininame{}.} \label{fig:MiniModelLayers}
    \label{fig:layer_versus_ndcg}
\vspace{-10pt}
\end{figure}

The results in Figure~\ref{fig:MiniModelLayers} show that there is a benefit beyond two layers and that at least four are required for the best performance. Performance stays the same with six, indicating that this might be the point of diminishing returns. Lastly, eight layers decrease performance. There could be a number of reasons for this, including that eight layers are harder to optimize or because effectively learning how to use all eight layers takes longer and thus might achieve better performance with extended training time.

\section{Conclusions}
We propose a new class of retrieval model—\name{}—that overcomes the limitations of similarity functions based on inner products that we prove to exist. Our model achieves a new state-of-the-art on TREC DL '19 and '20 for BERT sized encoder models with a single dense document vector and shows even stronger relative improvement on harder retrieval tasks such as tip-of-the-tongue queries. Additionally, we demonstrate that learned relevance models can be applied to large-scale search corpus in an efficient way with our proposed approximate search algorithm.
\medskip

{\footnotesize
\noindent {\textbf{Acknowledgments.}}
    This work was supported in part by the Center for Intelligent Information Retrieval, in part by the NSF Graduate Research Fellowships Program Award \#1938059, and in part by the Office of Naval Research contract number N000142412612. Any opinions, findings, conclusions or recommendations expressed in this material are those of the authors and do not necessarily reflect those of the sponsor.
}

\bibliographystyle{ACM-Reference-Format}
\balance
\bibliography{XX-references}


\begin{thebibliography}{81}

%%% the Web address.

\ifx \showCODEN    \undefined \def \showCODEN     #1{\unskip}     \fi
\ifx \showDOI      \undefined \def \showDOI       #1{#1}\fi
\ifx \showISBNx    \undefined \def \showISBNx     #1{\unskip}     \fi
\ifx \showISBNxiii \undefined \def \showISBNxiii  #1{\unskip}     \fi
\ifx \showISSN     \undefined \def \showISSN      #1{\unskip}     \fi
\ifx \showLCCN     \undefined \def \showLCCN      #1{\unskip}     \fi
\ifx \shownote     \undefined \def \shownote      #1{#1}          \fi
\ifx \showarticletitle \undefined \def \showarticletitle #1{#1}   \fi
\ifx \showURL      \undefined \def \showURL       {\relax}        \fi
% invisible to TeX
\providecommand\bibfield[2]{#2}
\providecommand\bibinfo[2]{#2}
\providecommand\natexlab[1]{#1}
\providecommand\showeprint[2][]{arXiv:#2}

\bibitem[Agarap(2018)]%
        {ReLU}
\bibfield{author}{\bibinfo{person}{Abien~Fred Agarap}.} \bibinfo{year}{2018}\natexlab{}.
\newblock \showarticletitle{Deep Learning using Rectified Linear Units (ReLU)}.
\newblock \bibinfo{journal}{\emph{CoRR}}  \bibinfo{volume}{abs/1803.08375} (\bibinfo{year}{2018}).
\newblock
\showeprint[arXiv]{1803.08375}
\urldef\tempurl%
\url{http://arxiv.org/abs/1803.08375}
\showURL{%
\tempurl}


\bibitem[Allan et~al\mbox{.}(2017)]%
        {TREC-Core17}
\bibfield{author}{\bibinfo{person}{James Allan}, \bibinfo{person}{Donna~K. Harman}, \bibinfo{person}{E. Kanoulas}, \bibinfo{person}{Dan Li}, \bibinfo{person}{Christophe~Van Gysel}, {and} \bibinfo{person}{Ellen~M. Voorhees}.} \bibinfo{year}{2017}\natexlab{}.
\newblock \showarticletitle{TREC 2017 Common Core Track Overview}. In \bibinfo{booktitle}{\emph{Text Retrieval Conference}}.
\newblock
\urldef\tempurl%
\url{https://api.semanticscholar.org/CorpusID:38019792}
\showURL{%
\tempurl}


\bibitem[Arguello et~al\mbox{.}(2023)]%
        {TREC-TOT-2023}
\bibfield{author}{\bibinfo{person}{Jaime Arguello}, \bibinfo{person}{Samarth Bhargav}, \bibinfo{person}{Fernando Diaz}, \bibinfo{person}{Evangelos Kanoulas}, {and} \bibinfo{person}{Bhaskar Mitra}.} \bibinfo{year}{2023}\natexlab{}.
\newblock \showarticletitle{Overview of the {TREC} 2023 Tip-of-the-Tongue Track}. In \bibinfo{booktitle}{\emph{The Thirty-Second Text REtrieval Conference Proceedings {(TREC} 2023), Gaithersburg, MD, USA, November 14-17, 2023}} \emph{(\bibinfo{series}{{NIST} Special Publication}, Vol.~\bibinfo{volume}{500-xxx})}, \bibfield{editor}{\bibinfo{person}{Ian Soboroff} {and} \bibinfo{person}{Angela Ellis}} (Eds.). \bibinfo{publisher}{National Institute of Standards and Technology {(NIST)}}.
\newblock
\urldef\tempurl%
\url{https://trec.nist.gov/pubs/trec32/papers/Overview\_tot.pdf}
\showURL{%
\tempurl}


\bibitem[Ba et~al\mbox{.}(2016)]%
        {LayerNorm}
\bibfield{author}{\bibinfo{person}{Lei~Jimmy Ba}, \bibinfo{person}{Jamie~Ryan Kiros}, {and} \bibinfo{person}{Geoffrey~E. Hinton}.} \bibinfo{year}{2016}\natexlab{}.
\newblock \showarticletitle{Layer Normalization}.
\newblock \bibinfo{journal}{\emph{CoRR}}  \bibinfo{volume}{abs/1607.06450} (\bibinfo{year}{2016}).
\newblock
\showeprint[arXiv]{1607.06450}
\urldef\tempurl%
\url{http://arxiv.org/abs/1607.06450}
\showURL{%
\tempurl}


\bibitem[Bhargav et~al\mbox{.}(2022)]%
        {TOMT-TrainingDataset}
\bibfield{author}{\bibinfo{person}{Samarth Bhargav}, \bibinfo{person}{Georgios Sidiropoulos}, {and} \bibinfo{person}{Evangelos Kanoulas}.} \bibinfo{year}{2022}\natexlab{}.
\newblock \showarticletitle{'It's on the tip of my tongue': {A} new Dataset for Known-Item Retrieval}. In \bibinfo{booktitle}{\emph{{WSDM} '22: The Fifteenth {ACM} International Conference on Web Search and Data Mining, Virtual Event / Tempe, AZ, USA, February 21 - 25, 2022}}, \bibfield{editor}{\bibinfo{person}{K.~Selcuk Candan}, \bibinfo{person}{Huan Liu}, \bibinfo{person}{Leman Akoglu}, \bibinfo{person}{Xin~Luna Dong}, {and} \bibinfo{person}{Jiliang Tang}} (Eds.). \bibinfo{publisher}{{ACM}}, \bibinfo{pages}{48--56}.
\newblock
\urldef\tempurl%
\url{https://doi.org/10.1145/3488560.3498421}
\showDOI{\tempurl}


\bibitem[Bondarenko et~al\mbox{.}(2020)]%
        {Touche}
\bibfield{author}{\bibinfo{person}{Alexander Bondarenko}, \bibinfo{person}{Maik Fr{\"{o}}be}, \bibinfo{person}{Meriem Beloucif}, \bibinfo{person}{Lukas Gienapp}, \bibinfo{person}{Yamen Ajjour}, \bibinfo{person}{Alexander Panchenko}, \bibinfo{person}{Chris Biemann}, \bibinfo{person}{Benno Stein}, \bibinfo{person}{Henning Wachsmuth}, \bibinfo{person}{Martin Potthast}, {and} \bibinfo{person}{Matthias Hagen}.} \bibinfo{year}{2020}\natexlab{}.
\newblock \showarticletitle{Overview of Touch{\'{e}} 2020: Argument Retrieval}. In \bibinfo{booktitle}{\emph{Working Notes of {CLEF} 2020 - Conference and Labs of the Evaluation Forum, Thessaloniki, Greece, September 22-25, 2020}} \emph{(\bibinfo{series}{{CEUR} Workshop Proceedings}, Vol.~\bibinfo{volume}{2696})}, \bibfield{editor}{\bibinfo{person}{Linda Cappellato}, \bibinfo{person}{Carsten Eickhoff}, \bibinfo{person}{Nicola Ferro}, {and} \bibinfo{person}{Aur{\'{e}}lie N{\'{e}}v{\'{e}}ol}} (Eds.). \bibinfo{publisher}{CEUR-WS.org}.
\newblock
\urldef\tempurl%
\url{https://ceur-ws.org/Vol-2696/paper\_261.pdf}
\showURL{%
\tempurl}


\bibitem[Boteva et~al\mbox{.}(2016)]%
        {NFCorpus}
\bibfield{author}{\bibinfo{person}{Vera Boteva}, \bibinfo{person}{Demian~Gholipour Ghalandari}, \bibinfo{person}{Artem Sokolov}, {and} \bibinfo{person}{Stefan Riezler}.} \bibinfo{year}{2016}\natexlab{}.
\newblock \showarticletitle{A Full-Text Learning to Rank Dataset for Medical Information Retrieval}. In \bibinfo{booktitle}{\emph{Advances in Information Retrieval - 38th European Conference on {IR} Research, {ECIR} 2016, Padua, Italy, March 20-23, 2016. Proceedings}} \emph{(\bibinfo{series}{Lecture Notes in Computer Science}, Vol.~\bibinfo{volume}{9626})}, \bibfield{editor}{\bibinfo{person}{Nicola Ferro}, \bibinfo{person}{Fabio Crestani}, \bibinfo{person}{Marie{-}Francine Moens}, \bibinfo{person}{Josiane Mothe}, \bibinfo{person}{Fabrizio Silvestri}, \bibinfo{person}{Giorgio Maria~Di Nunzio}, \bibinfo{person}{Claudia Hauff}, {and} \bibinfo{person}{Gianmaria Silvello}} (Eds.). \bibinfo{publisher}{Springer}, \bibinfo{pages}{716--722}.
\newblock
\urldef\tempurl%
\url{https://doi.org/10.1007/978-3-319-30671-1\_58}
\showDOI{\tempurl}


\bibitem[Burges et~al\mbox{.}(2006)]%
        {LearningToRankNonSmooth}
\bibfield{author}{\bibinfo{person}{Christopher Burges}, \bibinfo{person}{Robert Ragno}, {and} \bibinfo{person}{Quoc Le}.} \bibinfo{year}{2006}\natexlab{}.
\newblock \showarticletitle{Learning to rank with nonsmooth cost functions}.
\newblock \bibinfo{journal}{\emph{Advances in neural information processing systems}}  \bibinfo{volume}{19} (\bibinfo{year}{2006}).
\newblock


\bibitem[Burges et~al\mbox{.}(2005)]%
        {RankNet}
\bibfield{author}{\bibinfo{person}{Christopher J.~C. Burges}, \bibinfo{person}{Tal Shaked}, \bibinfo{person}{Erin Renshaw}, \bibinfo{person}{Ari Lazier}, \bibinfo{person}{Matt Deeds}, \bibinfo{person}{Nicole Hamilton}, {and} \bibinfo{person}{Gregory~N. Hullender}.} \bibinfo{year}{2005}\natexlab{}.
\newblock \showarticletitle{Learning to rank using gradient descent}. In \bibinfo{booktitle}{\emph{Machine Learning, Proceedings of the Twenty-Second International Conference {(ICML} 2005), Bonn, Germany, August 7-11, 2005}} \emph{(\bibinfo{series}{{ACM} International Conference Proceeding Series}, Vol.~\bibinfo{volume}{119})}, \bibfield{editor}{\bibinfo{person}{Luc~De Raedt} {and} \bibinfo{person}{Stefan Wrobel}} (Eds.). \bibinfo{publisher}{{ACM}}, \bibinfo{pages}{89--96}.
\newblock
\urldef\tempurl%
\url{https://doi.org/10.1145/1102351.1102363}
\showDOI{\tempurl}


\bibitem[Craswell et~al\mbox{.}(2020a)]%
        {TREC_DL_2020}
\bibfield{author}{\bibinfo{person}{Nick Craswell}, \bibinfo{person}{Bhaskar Mitra}, \bibinfo{person}{Emine Yilmaz}, {and} \bibinfo{person}{Daniel Campos}.} \bibinfo{year}{2020}\natexlab{a}.
\newblock \showarticletitle{Overview of the {TREC} 2020 Deep Learning Track}. In \bibinfo{booktitle}{\emph{Proceedings of the Twenty-Ninth Text REtrieval Conference, {TREC} 2020, Virtual Event [Gaithersburg, Maryland, USA], November 16-20, 2020}} \emph{(\bibinfo{series}{{NIST} Special Publication}, Vol.~\bibinfo{volume}{1266})}, \bibfield{editor}{\bibinfo{person}{Ellen~M. Voorhees} {and} \bibinfo{person}{Angela Ellis}} (Eds.). \bibinfo{publisher}{National Institute of Standards and Technology {(NIST)}}.
\newblock
\urldef\tempurl%
\url{https://trec.nist.gov/pubs/trec29/papers/OVERVIEW.DL.pdf}
\showURL{%
\tempurl}


\bibitem[Craswell et~al\mbox{.}(2021)]%
        {TREC-DL-21}
\bibfield{author}{\bibinfo{person}{Nick Craswell}, \bibinfo{person}{Bhaskar Mitra}, \bibinfo{person}{Emine Yilmaz}, \bibinfo{person}{Daniel Campos}, {and} \bibinfo{person}{Jimmy Lin}.} \bibinfo{year}{2021}\natexlab{}.
\newblock \showarticletitle{Overview of the {TREC} 2021 Deep Learning Track}. In \bibinfo{booktitle}{\emph{Proceedings of the Thirtieth Text REtrieval Conference, {TREC} 2021, online, November 15-19, 2021}} \emph{(\bibinfo{series}{{NIST} Special Publication}, Vol.~\bibinfo{volume}{500-335})}, \bibfield{editor}{\bibinfo{person}{Ian Soboroff} {and} \bibinfo{person}{Angela Ellis}} (Eds.). \bibinfo{publisher}{National Institute of Standards and Technology {(NIST)}}.
\newblock
\urldef\tempurl%
\url{https://trec.nist.gov/pubs/trec30/papers/Overview-DL.pdf}
\showURL{%
\tempurl}


\bibitem[Craswell et~al\mbox{.}(2022)]%
        {TREC-DL-22}
\bibfield{author}{\bibinfo{person}{Nick Craswell}, \bibinfo{person}{Bhaskar Mitra}, \bibinfo{person}{Emine Yilmaz}, \bibinfo{person}{Daniel Campos}, \bibinfo{person}{Jimmy Lin}, \bibinfo{person}{Ellen~M. Voorhees}, {and} \bibinfo{person}{Ian Soboroff}.} \bibinfo{year}{2022}\natexlab{}.
\newblock \showarticletitle{Overview of the {TREC} 2022 Deep Learning Track}. In \bibinfo{booktitle}{\emph{Proceedings of the Thirty-First Text REtrieval Conference, {TREC} 2022, online, November 15-19, 2022}} \emph{(\bibinfo{series}{{NIST} Special Publication}, Vol.~\bibinfo{volume}{500-338})}, \bibfield{editor}{\bibinfo{person}{Ian Soboroff} {and} \bibinfo{person}{Angela Ellis}} (Eds.). \bibinfo{publisher}{National Institute of Standards and Technology {(NIST)}}.
\newblock
\urldef\tempurl%
\url{https://trec.nist.gov/pubs/trec31/papers/Overview\_deep.pdf}
\showURL{%
\tempurl}


\bibitem[Craswell et~al\mbox{.}(2020b)]%
        {TREC_DL_2019}
\bibfield{author}{\bibinfo{person}{Nick Craswell}, \bibinfo{person}{Bhaskar Mitra}, \bibinfo{person}{Emine Yilmaz}, \bibinfo{person}{Daniel Campos}, {and} \bibinfo{person}{Ellen~M. Voorhees}.} \bibinfo{year}{2020}\natexlab{b}.
\newblock \showarticletitle{Overview of the {TREC} 2019 deep learning track}.
\newblock \bibinfo{journal}{\emph{CoRR}}  \bibinfo{volume}{abs/2003.07820} (\bibinfo{year}{2020}).
\newblock
\showeprint[arXiv]{2003.07820}
\urldef\tempurl%
\url{https://arxiv.org/abs/2003.07820}
\showURL{%
\tempurl}


\bibitem[Dehghani et~al\mbox{.}(2017)]%
        {Dehghani2017WeakSupervision}
\bibfield{author}{\bibinfo{person}{Mostafa Dehghani}, \bibinfo{person}{Hamed Zamani}, \bibinfo{person}{Aliaksei Severyn}, \bibinfo{person}{Jaap Kamps}, {and} \bibinfo{person}{W.~Bruce Croft}.} \bibinfo{year}{2017}\natexlab{}.
\newblock \showarticletitle{Neural Ranking Models with Weak Supervision}. In \bibinfo{booktitle}{\emph{Proceedings of the 40th International ACM SIGIR Conference on Research and Development in Information Retrieval}} (Shinjuku, Tokyo, Japan) \emph{(\bibinfo{series}{SIGIR '17})}. \bibinfo{publisher}{Association for Computing Machinery}, \bibinfo{address}{New York, NY, USA}, \bibinfo{pages}{65–74}.
\newblock
\showISBNx{9781450350228}
\urldef\tempurl%
\url{https://doi.org/10.1145/3077136.3080832}
\showDOI{\tempurl}


\bibitem[Del{\'{e}}tang et~al\mbox{.}(2024)]%
        {LanguageModelingIsCompression}
\bibfield{author}{\bibinfo{person}{Gr{\'{e}}goire Del{\'{e}}tang}, \bibinfo{person}{Anian Ruoss}, \bibinfo{person}{Paul{-}Ambroise Duquenne}, \bibinfo{person}{Elliot Catt}, \bibinfo{person}{Tim Genewein}, \bibinfo{person}{Christopher Mattern}, \bibinfo{person}{Jordi Grau{-}Moya}, \bibinfo{person}{Li~Kevin Wenliang}, \bibinfo{person}{Matthew Aitchison}, \bibinfo{person}{Laurent Orseau}, \bibinfo{person}{Marcus Hutter}, {and} \bibinfo{person}{Joel Veness}.} \bibinfo{year}{2024}\natexlab{}.
\newblock \showarticletitle{Language Modeling Is Compression}. In \bibinfo{booktitle}{\emph{The Twelfth International Conference on Learning Representations, {ICLR} 2024, Vienna, Austria, May 7-11, 2024}}. \bibinfo{publisher}{OpenReview.net}.
\newblock
\urldef\tempurl%
\url{https://openreview.net/forum?id=jznbgiynus}
\showURL{%
\tempurl}


\bibitem[Devlin et~al\mbox{.}(2019)]%
        {BERT}
\bibfield{author}{\bibinfo{person}{Jacob Devlin}, \bibinfo{person}{Ming{-}Wei Chang}, \bibinfo{person}{Kenton Lee}, {and} \bibinfo{person}{Kristina Toutanova}.} \bibinfo{year}{2019}\natexlab{}.
\newblock \showarticletitle{{BERT:} Pre-training of Deep Bidirectional Transformers for Language Understanding}. In \bibinfo{booktitle}{\emph{Proceedings of the 2019 Conference of the North American Chapter of the Association for Computational Linguistics: Human Language Technologies, {NAACL-HLT} 2019, Minneapolis, MN, USA, June 2-7, 2019, Volume 1 (Long and Short Papers)}}, \bibfield{editor}{\bibinfo{person}{Jill Burstein}, \bibinfo{person}{Christy Doran}, {and} \bibinfo{person}{Thamar Solorio}} (Eds.). \bibinfo{publisher}{Association for Computational Linguistics}, \bibinfo{pages}{4171--4186}.
\newblock
\urldef\tempurl%
\url{https://doi.org/10.18653/V1/N19-1423}
\showDOI{\tempurl}


\bibitem[Formal et~al\mbox{.}(2022)]%
        {SPLADE++}
\bibfield{author}{\bibinfo{person}{Thibault Formal}, \bibinfo{person}{Carlos Lassance}, \bibinfo{person}{Benjamin Piwowarski}, {and} \bibinfo{person}{St{\'{e}}phane Clinchant}.} \bibinfo{year}{2022}\natexlab{}.
\newblock \showarticletitle{From Distillation to Hard Negative Sampling: Making Sparse Neural {IR} Models More Effective}. In \bibinfo{booktitle}{\emph{{SIGIR} '22: The 45th International {ACM} {SIGIR} Conference on Research and Development in Information Retrieval, Madrid, Spain, July 11 - 15, 2022}}, \bibfield{editor}{\bibinfo{person}{Enrique Amig{\'{o}}}, \bibinfo{person}{Pablo Castells}, \bibinfo{person}{Julio Gonzalo}, \bibinfo{person}{Ben Carterette}, \bibinfo{person}{J.~Shane Culpepper}, {and} \bibinfo{person}{Gabriella Kazai}} (Eds.). \bibinfo{publisher}{{ACM}}, \bibinfo{pages}{2353--2359}.
\newblock
\urldef\tempurl%
\url{https://doi.org/10.1145/3477495.3531857}
\showDOI{\tempurl}


\bibitem[Formal et~al\mbox{.}(2021)]%
        {SPLADE}
\bibfield{author}{\bibinfo{person}{Thibault Formal}, \bibinfo{person}{Benjamin Piwowarski}, {and} \bibinfo{person}{St\'{e}phane Clinchant}.} \bibinfo{year}{2021}\natexlab{}.
\newblock \showarticletitle{SPLADE: Sparse Lexical and Expansion Model for First Stage Ranking}. In \bibinfo{booktitle}{\emph{Proceedings of the 44th International ACM SIGIR Conference on Research and Development in Information Retrieval}}. \bibinfo{publisher}{Association for Computing Machinery}, \bibinfo{address}{New York, NY, USA}, \bibinfo{pages}{2288–2292}.
\newblock
\showISBNx{9781450380379}
\urldef\tempurl%
\url{https://doi.org/10.1145/3404835.3463098}
\showURL{%
\tempurl}


\bibitem[Gao and Callan(2022)]%
        {CoCondensor}
\bibfield{author}{\bibinfo{person}{Luyu Gao} {and} \bibinfo{person}{Jamie Callan}.} \bibinfo{year}{2022}\natexlab{}.
\newblock \showarticletitle{Unsupervised Corpus Aware Language Model Pre-training for Dense Passage Retrieval}. In \bibinfo{booktitle}{\emph{Proceedings of the 60th Annual Meeting of the Association for Computational Linguistics (Volume 1: Long Papers), {ACL} 2022, Dublin, Ireland, May 22-27, 2022}}, \bibfield{editor}{\bibinfo{person}{Smaranda Muresan}, \bibinfo{person}{Preslav Nakov}, {and} \bibinfo{person}{Aline Villavicencio}} (Eds.). \bibinfo{publisher}{Association for Computational Linguistics}, \bibinfo{pages}{2843--2853}.
\newblock
\urldef\tempurl%
\url{https://doi.org/10.18653/V1/2022.ACL-LONG.203}
\showDOI{\tempurl}


\bibitem[Gao et~al\mbox{.}(2021)]%
        {coil}
\bibfield{author}{\bibinfo{person}{Luyu Gao}, \bibinfo{person}{Zhuyun Dai}, {and} \bibinfo{person}{Jamie Callan}.} \bibinfo{year}{2021}\natexlab{}.
\newblock \showarticletitle{COIL: Revisit Exact Lexical Match in Information Retrieval with Contextualized Inverted List}. In \bibinfo{booktitle}{\emph{North American Chapter of the Association for Computational Linguistics}}.
\newblock
\urldef\tempurl%
\url{https://api.semanticscholar.org/CorpusID:233241070}
\showURL{%
\tempurl}


\bibitem[Guo et~al\mbox{.}(2020)]%
        {Guo2020}
\bibfield{author}{\bibinfo{person}{Jiafeng Guo}, \bibinfo{person}{Yixing Fan}, \bibinfo{person}{Liang Pang}, \bibinfo{person}{Liu Yang}, \bibinfo{person}{Qingyao Ai}, \bibinfo{person}{Hamed Zamani}, \bibinfo{person}{Chen Wu}, \bibinfo{person}{W.~Bruce Croft}, {and} \bibinfo{person}{Xueqi Cheng}.} \bibinfo{year}{2020}\natexlab{}.
\newblock \showarticletitle{A Deep Look into neural ranking models for information retrieval}.
\newblock \bibinfo{journal}{\emph{Information Processing \& Management}} \bibinfo{volume}{57}, \bibinfo{number}{6} (\bibinfo{year}{2020}), \bibinfo{pages}{102067}.
\newblock
\showISSN{0306-4573}
\urldef\tempurl%
\url{https://doi.org/10.1016/j.ipm.2019.102067}
\showDOI{\tempurl}


\bibitem[Ha et~al\mbox{.}(2017)]%
        {HyperNetworks}
\bibfield{author}{\bibinfo{person}{David Ha}, \bibinfo{person}{Andrew~M. Dai}, {and} \bibinfo{person}{Quoc~V. Le}.} \bibinfo{year}{2017}\natexlab{}.
\newblock \showarticletitle{HyperNetworks}. In \bibinfo{booktitle}{\emph{5th International Conference on Learning Representations, {ICLR} 2017, Toulon, France, April 24-26, 2017, Conference Track Proceedings}}. \bibinfo{publisher}{OpenReview.net}.
\newblock
\urldef\tempurl%
\url{https://openreview.net/forum?id=rkpACe1lx}
\showURL{%
\tempurl}


\bibitem[Hasibi et~al\mbox{.}(2017)]%
        {DBpedia}
\bibfield{author}{\bibinfo{person}{Faegheh Hasibi}, \bibinfo{person}{Fedor Nikolaev}, \bibinfo{person}{Chenyan Xiong}, \bibinfo{person}{Krisztian Balog}, \bibinfo{person}{Svein~Erik Bratsberg}, \bibinfo{person}{Alexander Kotov}, {and} \bibinfo{person}{Jamie Callan}.} \bibinfo{year}{2017}\natexlab{}.
\newblock \showarticletitle{DBpedia-Entity v2: {A} Test Collection for Entity Search}. In \bibinfo{booktitle}{\emph{Proceedings of the 40th International {ACM} {SIGIR} Conference on Research and Development in Information Retrieval, Shinjuku, Tokyo, Japan, August 7-11, 2017}}, \bibfield{editor}{\bibinfo{person}{Noriko Kando}, \bibinfo{person}{Tetsuya Sakai}, \bibinfo{person}{Hideo Joho}, \bibinfo{person}{Hang Li}, \bibinfo{person}{Arjen~P. de~Vries}, {and} \bibinfo{person}{Ryen~W. White}} (Eds.). \bibinfo{publisher}{{ACM}}, \bibinfo{pages}{1265--1268}.
\newblock
\urldef\tempurl%
\url{https://doi.org/10.1145/3077136.3080751}
\showDOI{\tempurl}


\bibitem[He and Chua(2017)]%
        {NeuralFactorizationMachines}
\bibfield{author}{\bibinfo{person}{Xiangnan He} {and} \bibinfo{person}{Tat{-}Seng Chua}.} \bibinfo{year}{2017}\natexlab{}.
\newblock \showarticletitle{Neural Factorization Machines for Sparse Predictive Analytics}. In \bibinfo{booktitle}{\emph{Proceedings of the 40th International {ACM} {SIGIR} Conference on Research and Development in Information Retrieval, Shinjuku, Tokyo, Japan, August 7-11, 2017}}, \bibfield{editor}{\bibinfo{person}{Noriko Kando}, \bibinfo{person}{Tetsuya Sakai}, \bibinfo{person}{Hideo Joho}, \bibinfo{person}{Hang Li}, \bibinfo{person}{Arjen~P. de~Vries}, {and} \bibinfo{person}{Ryen~W. White}} (Eds.). \bibinfo{publisher}{{ACM}}, \bibinfo{pages}{355--364}.
\newblock
\urldef\tempurl%
\url{https://doi.org/10.1145/3077136.3080777}
\showDOI{\tempurl}


\bibitem[He et~al\mbox{.}(2017)]%
        {LearnedCollaborativeFiltering}
\bibfield{author}{\bibinfo{person}{Xiangnan He}, \bibinfo{person}{Lizi Liao}, \bibinfo{person}{Hanwang Zhang}, \bibinfo{person}{Liqiang Nie}, \bibinfo{person}{Xia Hu}, {and} \bibinfo{person}{Tat{-}Seng Chua}.} \bibinfo{year}{2017}\natexlab{}.
\newblock \showarticletitle{Neural Collaborative Filtering}. In \bibinfo{booktitle}{\emph{Proceedings of the 26th International Conference on World Wide Web, {WWW} 2017, Perth, Australia, April 3-7, 2017}}, \bibfield{editor}{\bibinfo{person}{Rick Barrett}, \bibinfo{person}{Rick Cummings}, \bibinfo{person}{Eugene Agichtein}, {and} \bibinfo{person}{Evgeniy Gabrilovich}} (Eds.). \bibinfo{publisher}{{ACM}}, \bibinfo{pages}{173--182}.
\newblock
\urldef\tempurl%
\url{https://doi.org/10.1145/3038912.3052569}
\showDOI{\tempurl}


\bibitem[Hofst{\"{a}}tter et~al\mbox{.}(2020)]%
        {MarginMSE}
\bibfield{author}{\bibinfo{person}{Sebastian Hofst{\"{a}}tter}, \bibinfo{person}{Sophia Althammer}, \bibinfo{person}{Michael Schr{\"{o}}der}, \bibinfo{person}{Mete Sertkan}, {and} \bibinfo{person}{Allan Hanbury}.} \bibinfo{year}{2020}\natexlab{}.
\newblock \showarticletitle{Improving Efficient Neural Ranking Models with Cross-Architecture Knowledge Distillation}.
\newblock \bibinfo{journal}{\emph{CoRR}}  \bibinfo{volume}{abs/2010.02666} (\bibinfo{year}{2020}).
\newblock
\showeprint[arXiv]{2010.02666}
\urldef\tempurl%
\url{https://arxiv.org/abs/2010.02666}
\showURL{%
\tempurl}


\bibitem[Hofst{\"a}tter et~al\mbox{.}(2022)]%
        {colberter}
\bibfield{author}{\bibinfo{person}{Sebastian Hofst{\"a}tter}, \bibinfo{person}{O. Khattab}, \bibinfo{person}{Sophia Althammer}, \bibinfo{person}{Mete Sertkan}, {and} \bibinfo{person}{Allan Hanbury}.} \bibinfo{year}{2022}\natexlab{}.
\newblock \showarticletitle{colberter}.
\newblock \bibinfo{journal}{\emph{Proceedings of the 31st ACM International Conference on Information \& Knowledge Management}} (\bibinfo{year}{2022}).
\newblock
\urldef\tempurl%
\url{https://api.semanticscholar.org/CorpusID:247628023}
\showURL{%
\tempurl}


\bibitem[Hofst{\"{a}}tter et~al\mbox{.}(2021)]%
        {TAS-B}
\bibfield{author}{\bibinfo{person}{Sebastian Hofst{\"{a}}tter}, \bibinfo{person}{Sheng{-}Chieh Lin}, \bibinfo{person}{Jheng{-}Hong Yang}, \bibinfo{person}{Jimmy Lin}, {and} \bibinfo{person}{Allan Hanbury}.} \bibinfo{year}{2021}\natexlab{}.
\newblock \showarticletitle{Efficiently Teaching an Effective Dense Retriever with Balanced Topic Aware Sampling}. In \bibinfo{booktitle}{\emph{{SIGIR} '21: The 44th International {ACM} {SIGIR} Conference on Research and Development in Information Retrieval, Virtual Event, Canada, July 11-15, 2021}}, \bibfield{editor}{\bibinfo{person}{Fernando Diaz}, \bibinfo{person}{Chirag Shah}, \bibinfo{person}{Torsten Suel}, \bibinfo{person}{Pablo Castells}, \bibinfo{person}{Rosie Jones}, {and} \bibinfo{person}{Tetsuya Sakai}} (Eds.). \bibinfo{publisher}{{ACM}}, \bibinfo{pages}{113--122}.
\newblock
\urldef\tempurl%
\url{https://doi.org/10.1145/3404835.3462891}
\showDOI{\tempurl}


\bibitem[Hornik et~al\mbox{.}(1989)]%
        {UniversalApproximators}
\bibfield{author}{\bibinfo{person}{Kurt Hornik}, \bibinfo{person}{Maxwell~B. Stinchcombe}, {and} \bibinfo{person}{Halbert White}.} \bibinfo{year}{1989}\natexlab{}.
\newblock \showarticletitle{Multilayer feedforward networks are universal approximators}.
\newblock \bibinfo{journal}{\emph{Neural Networks}} \bibinfo{volume}{2}, \bibinfo{number}{5} (\bibinfo{year}{1989}), \bibinfo{pages}{359--366}.
\newblock
\urldef\tempurl%
\url{https://doi.org/10.1016/0893-6080(89)90020-8}
\showDOI{\tempurl}


\bibitem[Izacard et~al\mbox{.}(2022)]%
        {Contriever}
\bibfield{author}{\bibinfo{person}{Gautier Izacard}, \bibinfo{person}{Mathilde Caron}, \bibinfo{person}{Lucas Hosseini}, \bibinfo{person}{Sebastian Riedel}, \bibinfo{person}{Piotr Bojanowski}, \bibinfo{person}{Armand Joulin}, {and} \bibinfo{person}{Edouard Grave}.} \bibinfo{year}{2022}\natexlab{}.
\newblock \showarticletitle{Unsupervised Dense Information Retrieval with Contrastive Learning}.
\newblock \bibinfo{journal}{\emph{Trans. Mach. Learn. Res.}}  \bibinfo{volume}{2022} (\bibinfo{year}{2022}).
\newblock
\urldef\tempurl%
\url{https://openreview.net/forum?id=jKN1pXi7b0}
\showURL{%
\tempurl}


\bibitem[Ji et~al\mbox{.}(2024)]%
        {LITE}
\bibfield{author}{\bibinfo{person}{Ziwei Ji}, \bibinfo{person}{Himanshu Jain}, \bibinfo{person}{Andreas Veit}, \bibinfo{person}{Sashank~J. Reddi}, \bibinfo{person}{Sadeep Jayasumana}, \bibinfo{person}{Ankit~Singh Rawat}, \bibinfo{person}{Aditya~Krishna Menon}, \bibinfo{person}{Felix Yu}, {and} \bibinfo{person}{Sanjiv Kumar}.} \bibinfo{year}{2024}\natexlab{}.
\newblock \showarticletitle{Efficient Document Ranking with Learnable Late Interactions}.
\newblock \bibinfo{journal}{\emph{CoRR}}  \bibinfo{volume}{abs/2406.17968} (\bibinfo{year}{2024}).
\newblock
\urldef\tempurl%
\url{https://doi.org/10.48550/ARXIV.2406.17968}
\showDOI{\tempurl}
\showeprint[arXiv]{2406.17968}


\bibitem[Karpukhin et~al\mbox{.}(2020)]%
        {DPR}
\bibfield{author}{\bibinfo{person}{Vladimir Karpukhin}, \bibinfo{person}{Barlas Oguz}, \bibinfo{person}{Sewon Min}, \bibinfo{person}{Patrick S.~H. Lewis}, \bibinfo{person}{Ledell Wu}, \bibinfo{person}{Sergey Edunov}, \bibinfo{person}{Danqi Chen}, {and} \bibinfo{person}{Wen{-}tau Yih}.} \bibinfo{year}{2020}\natexlab{}.
\newblock \showarticletitle{Dense Passage Retrieval for Open-Domain Question Answering}. In \bibinfo{booktitle}{\emph{Proceedings of the 2020 Conference on Empirical Methods in Natural Language Processing, {EMNLP} 2020, Online, November 16-20, 2020}}, \bibfield{editor}{\bibinfo{person}{Bonnie Webber}, \bibinfo{person}{Trevor Cohn}, \bibinfo{person}{Yulan He}, {and} \bibinfo{person}{Yang Liu}} (Eds.). \bibinfo{publisher}{Association for Computational Linguistics}, \bibinfo{pages}{6769--6781}.
\newblock
\urldef\tempurl%
\url{https://doi.org/10.18653/V1/2020.EMNLP-MAIN.550}
\showDOI{\tempurl}


\bibitem[Khattab and Zaharia(2020)]%
        {ColBERT_v1}
\bibfield{author}{\bibinfo{person}{Omar Khattab} {and} \bibinfo{person}{Matei Zaharia}.} \bibinfo{year}{2020}\natexlab{}.
\newblock \showarticletitle{ColBERT: Efficient and Effective Passage Search via Contextualized Late Interaction over {BERT}}. In \bibinfo{booktitle}{\emph{Proceedings of the 43rd International {ACM} {SIGIR} conference on research and development in Information Retrieval, {SIGIR} 2020, Virtual Event, China, July 25-30, 2020}}, \bibfield{editor}{\bibinfo{person}{Jimmy~X. Huang}, \bibinfo{person}{Yi~Chang}, \bibinfo{person}{Xueqi Cheng}, \bibinfo{person}{Jaap Kamps}, \bibinfo{person}{Vanessa Murdock}, \bibinfo{person}{Ji{-}Rong Wen}, {and} \bibinfo{person}{Yiqun Liu}} (Eds.). \bibinfo{publisher}{{ACM}}, \bibinfo{pages}{39--48}.
\newblock
\urldef\tempurl%
\url{https://doi.org/10.1145/3397271.3401075}
\showDOI{\tempurl}


\bibitem[Li et~al\mbox{.}(2022)]%
        {citadel}
\bibfield{author}{\bibinfo{person}{Minghan Li}, \bibinfo{person}{Sheng-Chieh Lin}, \bibinfo{person}{Barlas Oğuz}, \bibinfo{person}{Asish Ghoshal}, \bibinfo{person}{Jimmy~J. Lin}, \bibinfo{person}{Yashar Mehdad}, \bibinfo{person}{Wen tau Yih}, {and} \bibinfo{person}{Xilun Chen}.} \bibinfo{year}{2022}\natexlab{}.
\newblock \showarticletitle{CITADEL: Conditional Token Interaction via Dynamic Lexical Routing for Efficient and Effective Multi-Vector Retrieval}. In \bibinfo{booktitle}{\emph{Annual Meeting of the Association for Computational Linguistics}}.
\newblock
\urldef\tempurl%
\url{https://api.semanticscholar.org/CorpusID:253708231}
\showURL{%
\tempurl}


\bibitem[Lifshits and Zhang(2009)]%
        {SmallWorld}
\bibfield{author}{\bibinfo{person}{Yury Lifshits} {and} \bibinfo{person}{Shengyu Zhang}.} \bibinfo{year}{2009}\natexlab{}.
\newblock \showarticletitle{Combinatorial algorithms for nearest neighbors, near-duplicates and small-world design}. In \bibinfo{booktitle}{\emph{Proceedings of the Twentieth Annual {ACM-SIAM} Symposium on Discrete Algorithms, {SODA} 2009, New York, NY, USA, January 4-6, 2009}}, \bibfield{editor}{\bibinfo{person}{Claire Mathieu}} (Ed.). \bibinfo{publisher}{{SIAM}}, \bibinfo{pages}{318--326}.
\newblock
\urldef\tempurl%
\url{https://doi.org/10.1137/1.9781611973068.36}
\showDOI{\tempurl}


\bibitem[Lin et~al\mbox{.}(2022)]%
        {lin2022pretrained}
\bibfield{author}{\bibinfo{person}{J. Lin}, \bibinfo{person}{R. Nogueira}, {and} \bibinfo{person}{A. Yates}.} \bibinfo{year}{2022}\natexlab{}.
\newblock \bibinfo{booktitle}{\emph{Pretrained Transformers for Text Ranking: BERT and Beyond}}.
\newblock \bibinfo{publisher}{Springer International Publishing}.
\newblock
\showISBNx{9783031021817}


\bibitem[Lin et~al\mbox{.}(2020)]%
        {TCT-ColBERT}
\bibfield{author}{\bibinfo{person}{Sheng{-}Chieh Lin}, \bibinfo{person}{Jheng{-}Hong Yang}, {and} \bibinfo{person}{Jimmy Lin}.} \bibinfo{year}{2020}\natexlab{}.
\newblock \showarticletitle{Distilling Dense Representations for Ranking using Tightly-Coupled Teachers}.
\newblock \bibinfo{journal}{\emph{CoRR}}  \bibinfo{volume}{abs/2010.11386} (\bibinfo{year}{2020}).
\newblock
\showeprint[arXiv]{2010.11386}
\urldef\tempurl%
\url{https://arxiv.org/abs/2010.11386}
\showURL{%
\tempurl}


\bibitem[Lin et~al\mbox{.}(2023a)]%
        {DRAGON}
\bibfield{author}{\bibinfo{person}{Sheng-Chieh Lin}, \bibinfo{person}{Akari Asai}, \bibinfo{person}{Minghan Li}, \bibinfo{person}{Barlas Oguz}, \bibinfo{person}{Jimmy Lin}, \bibinfo{person}{Yashar Mehdad}, \bibinfo{person}{Wen-tau Yih}, {and} \bibinfo{person}{Xilun Chen}.} \bibinfo{year}{2023}\natexlab{a}.
\newblock \showarticletitle{How to Train Your Dragon: Diverse Augmentation Towards Generalizable Dense Retrieval}. In \bibinfo{booktitle}{\emph{Findings of the Association for Computational Linguistics: EMNLP 2023}}, \bibfield{editor}{\bibinfo{person}{Houda Bouamor}, \bibinfo{person}{Juan Pino}, {and} \bibinfo{person}{Kalika Bali}} (Eds.). \bibinfo{publisher}{Association for Computational Linguistics}, \bibinfo{address}{Singapore}, \bibinfo{pages}{6385--6400}.
\newblock
\urldef\tempurl%
\url{https://doi.org/10.18653/v1/2023.findings-emnlp.423}
\showDOI{\tempurl}


\bibitem[Lin et~al\mbox{.}(2023b)]%
        {prod}
\bibfield{author}{\bibinfo{person}{Zhenghao Lin}, \bibinfo{person}{Yeyun Gong}, \bibinfo{person}{Xiao Liu}, \bibinfo{person}{Hang Zhang}, \bibinfo{person}{Chen Lin}, \bibinfo{person}{Anlei Dong}, \bibinfo{person}{Jian Jiao}, \bibinfo{person}{Jingwen Lu}, \bibinfo{person}{Daxin Jiang}, \bibinfo{person}{Rangan Majumder}, {and} \bibinfo{person}{Nan Duan}.} \bibinfo{year}{2023}\natexlab{b}.
\newblock \showarticletitle{PROD: Progressive Distillation for Dense Retrieval}. In \bibinfo{booktitle}{\emph{Proceedings of the ACM Web Conference 2023}} (Austin, TX, USA) \emph{(\bibinfo{series}{WWW '23})}. \bibinfo{publisher}{Association for Computing Machinery}, \bibinfo{address}{New York, NY, USA}, \bibinfo{pages}{3299–3308}.
\newblock
\showISBNx{9781450394161}
\urldef\tempurl%
\url{https://doi.org/10.1145/3543507.3583421}
\showDOI{\tempurl}


\bibitem[Ma et~al\mbox{.}(2021)]%
        {b-prop}
\bibfield{author}{\bibinfo{person}{Xinyu Ma}, \bibinfo{person}{Jiafeng Guo}, \bibinfo{person}{Ruqing Zhang}, \bibinfo{person}{Yixing Fan}, \bibinfo{person}{Yingyan Li}, {and} \bibinfo{person}{Xueqi Cheng}.} \bibinfo{year}{2021}\natexlab{}.
\newblock \showarticletitle{B-PROP: Bootstrapped Pre-training with Representative Words Prediction for Ad-hoc Retrieval}.
\newblock \bibinfo{journal}{\emph{Proceedings of the 44th International ACM SIGIR Conference on Research and Development in Information Retrieval}} (\bibinfo{year}{2021}).
\newblock
\urldef\tempurl%
\url{https://api.semanticscholar.org/CorpusID:233307194}
\showURL{%
\tempurl}


\bibitem[Ma et~al\mbox{.}(2024)]%
        {RepLlama}
\bibfield{author}{\bibinfo{person}{Xueguang Ma}, \bibinfo{person}{Liang Wang}, \bibinfo{person}{Nan Yang}, \bibinfo{person}{Furu Wei}, {and} \bibinfo{person}{Jimmy Lin}.} \bibinfo{year}{2024}\natexlab{}.
\newblock \showarticletitle{Fine-Tuning LLaMA for Multi-Stage Text Retrieval}. In \bibinfo{booktitle}{\emph{Proceedings of the 47th International {ACM} {SIGIR} Conference on Research and Development in Information Retrieval, {SIGIR} 2024, Washington DC, USA, July 14-18, 2024}}, \bibfield{editor}{\bibinfo{person}{Grace~Hui Yang}, \bibinfo{person}{Hongning Wang}, \bibinfo{person}{Sam Han}, \bibinfo{person}{Claudia Hauff}, \bibinfo{person}{Guido Zuccon}, {and} \bibinfo{person}{Yi~Zhang}} (Eds.). \bibinfo{publisher}{{ACM}}, \bibinfo{pages}{2421--2425}.
\newblock
\urldef\tempurl%
\url{https://doi.org/10.1145/3626772.3657951}
\showDOI{\tempurl}


\bibitem[MacAvaney et~al\mbox{.}(2020)]%
        {PreTTR}
\bibfield{author}{\bibinfo{person}{Sean MacAvaney}, \bibinfo{person}{Franco~Maria Nardini}, \bibinfo{person}{Raffaele Perego}, \bibinfo{person}{Nicola Tonellotto}, \bibinfo{person}{Nazli Goharian}, {and} \bibinfo{person}{Ophir Frieder}.} \bibinfo{year}{2020}\natexlab{}.
\newblock \showarticletitle{Efficient Document Re-Ranking for Transformers by Precomputing Term Representations}. In \bibinfo{booktitle}{\emph{Proceedings of the 43rd International {ACM} {SIGIR} conference on research and development in Information Retrieval, {SIGIR} 2020, Virtual Event, China, July 25-30, 2020}}, \bibfield{editor}{\bibinfo{person}{Jimmy~X. Huang}, \bibinfo{person}{Yi~Chang}, \bibinfo{person}{Xueqi Cheng}, \bibinfo{person}{Jaap Kamps}, \bibinfo{person}{Vanessa Murdock}, \bibinfo{person}{Ji{-}Rong Wen}, {and} \bibinfo{person}{Yiqun Liu}} (Eds.). \bibinfo{publisher}{{ACM}}, \bibinfo{pages}{49--58}.
\newblock
\urldef\tempurl%
\url{https://doi.org/10.1145/3397271.3401093}
\showDOI{\tempurl}


\bibitem[Mackie et~al\mbox{.}(2021)]%
        {TREC_DL_HARD}
\bibfield{author}{\bibinfo{person}{Iain Mackie}, \bibinfo{person}{Jeffrey Dalton}, {and} \bibinfo{person}{Andrew Yates}.} \bibinfo{year}{2021}\natexlab{}.
\newblock \showarticletitle{How Deep is your Learning: the DL-HARD Annotated Deep Learning Dataset}. In \bibinfo{booktitle}{\emph{Proceedings of the 44th International ACM SIGIR Conference on Research and Development in Information Retrieval}}.
\newblock


\bibitem[Maia et~al\mbox{.}(2018)]%
        {FiQA}
\bibfield{author}{\bibinfo{person}{Macedo Maia}, \bibinfo{person}{Siegfried Handschuh}, \bibinfo{person}{Andr{\'{e}} Freitas}, \bibinfo{person}{Brian Davis}, \bibinfo{person}{Ross McDermott}, \bibinfo{person}{Manel Zarrouk}, {and} \bibinfo{person}{Alexandra Balahur}.} \bibinfo{year}{2018}\natexlab{}.
\newblock \showarticletitle{WWW'18 Open Challenge: Financial Opinion Mining and Question Answering}. In \bibinfo{booktitle}{\emph{Companion of the The Web Conference 2018 on The Web Conference 2018, {WWW} 2018, Lyon , France, April 23-27, 2018}}, \bibfield{editor}{\bibinfo{person}{Pierre{-}Antoine Champin}, \bibinfo{person}{Fabien Gandon}, \bibinfo{person}{Mounia Lalmas}, {and} \bibinfo{person}{Panagiotis~G. Ipeirotis}} (Eds.). \bibinfo{publisher}{{ACM}}, \bibinfo{pages}{1941--1942}.
\newblock
\urldef\tempurl%
\url{https://doi.org/10.1145/3184558.3192301}
\showDOI{\tempurl}


\bibitem[Malkov and Yashunin(2020)]%
        {HNSW}
\bibfield{author}{\bibinfo{person}{Yury~A. Malkov} {and} \bibinfo{person}{Dmitry~A. Yashunin}.} \bibinfo{year}{2020}\natexlab{}.
\newblock \showarticletitle{Efficient and Robust Approximate Nearest Neighbor Search Using Hierarchical Navigable Small World Graphs}.
\newblock \bibinfo{journal}{\emph{{IEEE} Trans. Pattern Anal. Mach. Intell.}} \bibinfo{volume}{42}, \bibinfo{number}{4} (\bibinfo{year}{2020}), \bibinfo{pages}{824--836}.
\newblock
\urldef\tempurl%
\url{https://doi.org/10.1109/TPAMI.2018.2889473}
\showDOI{\tempurl}


\bibitem[Mitra and Craswell(2018)]%
        {Mitra:2018:NeuralIR}
\bibfield{author}{\bibinfo{person}{Bhaskar Mitra} {and} \bibinfo{person}{Nick Craswell}.} \bibinfo{year}{2018}\natexlab{}.
\newblock \showarticletitle{An Introduction to Neural Information Retrieval}.
\newblock \bibinfo{journal}{\emph{Found. Trends Inf. Retr.}} \bibinfo{volume}{13}, \bibinfo{number}{1} (\bibinfo{date}{Dec.} \bibinfo{year}{2018}), \bibinfo{pages}{1–126}.
\newblock
\showISSN{1554-0669}
\urldef\tempurl%
\url{https://doi.org/10.1561/1500000061}
\showDOI{\tempurl}


\bibitem[Mitra and Craswell(2019)]%
        {DuetV2}
\bibfield{author}{\bibinfo{person}{Bhaskar Mitra} {and} \bibinfo{person}{Nick Craswell}.} \bibinfo{year}{2019}\natexlab{}.
\newblock \showarticletitle{An Updated Duet Model for Passage Re-ranking}.
\newblock \bibinfo{journal}{\emph{CoRR}}  \bibinfo{volume}{abs/1903.07666} (\bibinfo{year}{2019}).
\newblock
\showeprint[arXiv]{1903.07666}
\urldef\tempurl%
\url{http://arxiv.org/abs/1903.07666}
\showURL{%
\tempurl}


\bibitem[Mitra et~al\mbox{.}(2017)]%
        {Duetv1}
\bibfield{author}{\bibinfo{person}{Bhaskar Mitra}, \bibinfo{person}{Fernando Diaz}, {and} \bibinfo{person}{Nick Craswell}.} \bibinfo{year}{2017}\natexlab{}.
\newblock \showarticletitle{Learning to Match using Local and Distributed Representations of Text for Web Search}. In \bibinfo{booktitle}{\emph{Proceedings of the 26th International Conference on World Wide Web, {WWW} 2017, Perth, Australia, April 3-7, 2017}}, \bibfield{editor}{\bibinfo{person}{Rick Barrett}, \bibinfo{person}{Rick Cummings}, \bibinfo{person}{Eugene Agichtein}, {and} \bibinfo{person}{Evgeniy Gabrilovich}} (Eds.). \bibinfo{publisher}{{ACM}}, \bibinfo{pages}{1291--1299}.
\newblock
\urldef\tempurl%
\url{https://doi.org/10.1145/3038912.3052579}
\showDOI{\tempurl}


\bibitem[Nguyen et~al\mbox{.}(2016)]%
        {MSMARCO}
\bibfield{author}{\bibinfo{person}{Tri Nguyen}, \bibinfo{person}{Mir Rosenberg}, \bibinfo{person}{Xia Song}, \bibinfo{person}{Jianfeng Gao}, \bibinfo{person}{Saurabh Tiwary}, \bibinfo{person}{Rangan Majumder}, {and} \bibinfo{person}{Li Deng}.} \bibinfo{year}{2016}\natexlab{}.
\newblock \showarticletitle{{MS} {MARCO:} {A} Human Generated MAchine Reading COmprehension Dataset}. In \bibinfo{booktitle}{\emph{Proceedings of the Workshop on Cognitive Computation: Integrating neural and symbolic approaches 2016 co-located with the 30th Annual Conference on Neural Information Processing Systems {(NIPS} 2016), Barcelona, Spain, December 9, 2016}} \emph{(\bibinfo{series}{{CEUR} Workshop Proceedings}, Vol.~\bibinfo{volume}{1773})}, \bibfield{editor}{\bibinfo{person}{Tarek~Richard Besold}, \bibinfo{person}{Antoine Bordes}, \bibinfo{person}{Artur~S. d'Avila Garcez}, {and} \bibinfo{person}{Greg Wayne}} (Eds.). \bibinfo{publisher}{CEUR-WS.org}.
\newblock
\urldef\tempurl%
\url{https://ceur-ws.org/Vol-1773/CoCoNIPS\_2016\_paper9.pdf}
\showURL{%
\tempurl}


\bibitem[Nogueira et~al\mbox{.}(2019)]%
        {MonoBERT}
\bibfield{author}{\bibinfo{person}{Rodrigo~Frassetto Nogueira}, \bibinfo{person}{Wei Yang}, \bibinfo{person}{Kyunghyun Cho}, {and} \bibinfo{person}{Jimmy Lin}.} \bibinfo{year}{2019}\natexlab{}.
\newblock \showarticletitle{Multi-Stage Document Ranking with {BERT}}.
\newblock \bibinfo{journal}{\emph{CoRR}}  \bibinfo{volume}{abs/1910.14424} (\bibinfo{year}{2019}).
\newblock
\showeprint[arXiv]{1910.14424}
\urldef\tempurl%
\url{http://arxiv.org/abs/1910.14424}
\showURL{%
\tempurl}


\bibitem[Pang et~al\mbox{.}(2016)]%
        {MatchPyramid}
\bibfield{author}{\bibinfo{person}{Liang Pang}, \bibinfo{person}{Yanyan Lan}, \bibinfo{person}{Jiafeng Guo}, \bibinfo{person}{Jun Xu}, \bibinfo{person}{Shengxian Wan}, {and} \bibinfo{person}{Xueqi Cheng}.} \bibinfo{year}{2016}\natexlab{}.
\newblock \showarticletitle{Text Matching as Image Recognition}. In \bibinfo{booktitle}{\emph{Proceedings of the Thirtieth {AAAI} Conference on Artificial Intelligence, February 12-17, 2016, Phoenix, Arizona, {USA}}}, \bibfield{editor}{\bibinfo{person}{Dale Schuurmans} {and} \bibinfo{person}{Michael~P. Wellman}} (Eds.). \bibinfo{publisher}{{AAAI} Press}, \bibinfo{pages}{2793--2799}.
\newblock
\urldef\tempurl%
\url{https://doi.org/10.1609/AAAI.V30I1.10341}
\showDOI{\tempurl}


\bibitem[Prakash et~al\mbox{.}(2021)]%
        {RANCE}
\bibfield{author}{\bibinfo{person}{Prafull Prakash}, \bibinfo{person}{Julian Killingback}, {and} \bibinfo{person}{Hamed Zamani}.} \bibinfo{year}{2021}\natexlab{}.
\newblock \showarticletitle{Learning Robust Dense Retrieval Models from Incomplete Relevance Labels}. In \bibinfo{booktitle}{\emph{{SIGIR} '21: The 44th International {ACM} {SIGIR} Conference on Research and Development in Information Retrieval, Virtual Event, Canada, July 11-15, 2021}}, \bibfield{editor}{\bibinfo{person}{Fernando Diaz}, \bibinfo{person}{Chirag Shah}, \bibinfo{person}{Torsten Suel}, \bibinfo{person}{Pablo Castells}, \bibinfo{person}{Rosie Jones}, {and} \bibinfo{person}{Tetsuya Sakai}} (Eds.). \bibinfo{publisher}{{ACM}}, \bibinfo{pages}{1728--1732}.
\newblock
\urldef\tempurl%
\url{https://doi.org/10.1145/3404835.3463106}
\showDOI{\tempurl}


\bibitem[Qu et~al\mbox{.}(2021)]%
        {RocketQA}
\bibfield{author}{\bibinfo{person}{Yingqi Qu}, \bibinfo{person}{Yuchen Ding}, \bibinfo{person}{Jing Liu}, \bibinfo{person}{Kai Liu}, \bibinfo{person}{Ruiyang Ren}, \bibinfo{person}{Wayne~Xin Zhao}, \bibinfo{person}{Daxiang Dong}, \bibinfo{person}{Hua Wu}, {and} \bibinfo{person}{Haifeng Wang}.} \bibinfo{year}{2021}\natexlab{}.
\newblock \showarticletitle{RocketQA: An Optimized Training Approach to Dense Passage Retrieval for Open-Domain Question Answering}. In \bibinfo{booktitle}{\emph{Proceedings of the 2021 Conference of the North American Chapter of the Association for Computational Linguistics: Human Language Technologies, {NAACL-HLT} 2021, Online, June 6-11, 2021}}, \bibfield{editor}{\bibinfo{person}{Kristina Toutanova}, \bibinfo{person}{Anna Rumshisky}, \bibinfo{person}{Luke Zettlemoyer}, \bibinfo{person}{Dilek Hakkani{-}T{\"{u}}r}, \bibinfo{person}{Iz~Beltagy}, \bibinfo{person}{Steven Bethard}, \bibinfo{person}{Ryan Cotterell}, \bibinfo{person}{Tanmoy Chakraborty}, {and} \bibinfo{person}{Yichao Zhou}} (Eds.). \bibinfo{publisher}{Association for Computational Linguistics}, \bibinfo{pages}{5835--5847}.
\newblock
\urldef\tempurl%
\url{https://doi.org/10.18653/V1/2021.NAACL-MAIN.466}
\showDOI{\tempurl}


\bibitem[Radon(1921)]%
        {RadonTheorem}
\bibfield{author}{\bibinfo{person}{Johann Radon}.} \bibinfo{year}{1921}\natexlab{}.
\newblock \showarticletitle{Mengen konvexer Körper, die einen gemeinsamen Punkt enthalten}.
\newblock \bibinfo{journal}{\emph{Math. Ann.}} \bibinfo{volume}{83}, \bibinfo{number}{1} (\bibinfo{year}{1921}).
\newblock
\urldef\tempurl%
\url{https://doi.org/10.1007/BF01464231}
\showDOI{\tempurl}


\bibitem[Reimers and Gurevych(2019)]%
        {SentenceBERT}
\bibfield{author}{\bibinfo{person}{Nils Reimers} {and} \bibinfo{person}{Iryna Gurevych}.} \bibinfo{year}{2019}\natexlab{}.
\newblock \showarticletitle{Sentence-BERT: Sentence Embeddings using Siamese BERT-Networks}. In \bibinfo{booktitle}{\emph{Proceedings of the 2019 Conference on Empirical Methods in Natural Language Processing}}. \bibinfo{publisher}{Association for Computational Linguistics}.
\newblock
\urldef\tempurl%
\url{https://arxiv.org/abs/1908.10084}
\showURL{%
\tempurl}


\bibitem[Roberts et~al\mbox{.}(2021)]%
        {TREC-Covid}
\bibfield{author}{\bibinfo{person}{Kirk Roberts}, \bibinfo{person}{Tasmeer Alam}, \bibinfo{person}{Steven Bedrick}, \bibinfo{person}{Dina Demner{-}Fushman}, \bibinfo{person}{Kyle Lo}, \bibinfo{person}{Ian Soboroff}, \bibinfo{person}{Ellen~M. Voorhees}, \bibinfo{person}{Lucy~Lu Wang}, {and} \bibinfo{person}{William~R. Hersh}.} \bibinfo{year}{2021}\natexlab{}.
\newblock \showarticletitle{Searching for scientific evidence in a pandemic: An overview of {TREC-COVID}}.
\newblock \bibinfo{journal}{\emph{J. Biomed. Informatics}}  \bibinfo{volume}{121} (\bibinfo{year}{2021}), \bibinfo{pages}{103865}.
\newblock
\urldef\tempurl%
\url{https://doi.org/10.1016/J.JBI.2021.103865}
\showDOI{\tempurl}


\bibitem[Rusu et~al\mbox{.}(2019)]%
        {MetaLearningWithLatentEmbeddingOptimization}
\bibfield{author}{\bibinfo{person}{Andrei~A. Rusu}, \bibinfo{person}{Dushyant Rao}, \bibinfo{person}{Jakub Sygnowski}, \bibinfo{person}{Oriol Vinyals}, \bibinfo{person}{Razvan Pascanu}, \bibinfo{person}{Simon Osindero}, {and} \bibinfo{person}{Raia Hadsell}.} \bibinfo{year}{2019}\natexlab{}.
\newblock \showarticletitle{Meta-Learning with Latent Embedding Optimization}. In \bibinfo{booktitle}{\emph{7th International Conference on Learning Representations, {ICLR} 2019, New Orleans, LA, USA, May 6-9, 2019}}. \bibinfo{publisher}{OpenReview.net}.
\newblock
\urldef\tempurl%
\url{https://openreview.net/forum?id=BJgklhAcK7}
\showURL{%
\tempurl}


\bibitem[Santhanam et~al\mbox{.}(2022)]%
        {colbert_v2}
\bibfield{author}{\bibinfo{person}{Keshav Santhanam}, \bibinfo{person}{Omar Khattab}, \bibinfo{person}{Jon Saad-Falcon}, \bibinfo{person}{Christopher Potts}, {and} \bibinfo{person}{Matei Zaharia}.} \bibinfo{year}{2022}\natexlab{}.
\newblock \showarticletitle{{C}ol{BERT}v2: Effective and Efficient Retrieval via Lightweight Late Interaction}. In \bibinfo{booktitle}{\emph{Proceedings of the 2022 Conference of the North American Chapter of the Association for Computational Linguistics: Human Language Technologies}}, \bibfield{editor}{\bibinfo{person}{Marine Carpuat}, \bibinfo{person}{Marie-Catherine de~Marneffe}, {and} \bibinfo{person}{Ivan~Vladimir Meza~Ruiz}} (Eds.). \bibinfo{publisher}{Association for Computational Linguistics}, \bibinfo{address}{Seattle, United States}, \bibinfo{pages}{3715--3734}.
\newblock
\urldef\tempurl%
\url{https://doi.org/10.18653/v1/2022.naacl-main.272}
\showDOI{\tempurl}


\bibitem[Sendera et~al\mbox{.}(2023)]%
        {HyperShot}
\bibfield{author}{\bibinfo{person}{Marcin Sendera}, \bibinfo{person}{Marcin Przewiezlikowski}, \bibinfo{person}{Konrad Karanowski}, \bibinfo{person}{Maciej Zieba}, \bibinfo{person}{Jacek Tabor}, {and} \bibinfo{person}{Przemyslaw Spurek}.} \bibinfo{year}{2023}\natexlab{}.
\newblock \showarticletitle{HyperShot: Few-Shot Learning by Kernel HyperNetworks}. In \bibinfo{booktitle}{\emph{{IEEE/CVF} Winter Conference on Applications of Computer Vision, {WACV} 2023, Waikoloa, HI, USA, January 2-7, 2023}}. \bibinfo{publisher}{{IEEE}}, \bibinfo{pages}{2468--2477}.
\newblock
\urldef\tempurl%
\url{https://doi.org/10.1109/WACV56688.2023.00250}
\showDOI{\tempurl}


\bibitem[Soboroff et~al\mbox{.}(2020)]%
        {TREC-News20}
\bibfield{author}{\bibinfo{person}{Ian Soboroff}, \bibinfo{person}{Shudong Huang}, {and} \bibinfo{person}{Donna Harman}.} \bibinfo{year}{2020}\natexlab{}.
\newblock \showarticletitle{{TREC} 2020 News Track Overview}. In \bibinfo{booktitle}{\emph{Proceedings of the Twenty-Ninth Text REtrieval Conference, {TREC} 2020, Virtual Event [Gaithersburg, Maryland, USA], November 16-20, 2020}} \emph{(\bibinfo{series}{{NIST} Special Publication}, Vol.~\bibinfo{volume}{1266})}, \bibfield{editor}{\bibinfo{person}{Ellen~M. Voorhees} {and} \bibinfo{person}{Angela Ellis}} (Eds.). \bibinfo{publisher}{National Institute of Standards and Technology {(NIST)}}.
\newblock
\urldef\tempurl%
\url{https://trec.nist.gov/pubs/trec29/papers/OVERVIEW.N.pdf}
\showURL{%
\tempurl}


\bibitem[Tan et~al\mbox{.}(2021)]%
        {FastNeuralRankingOnBipartiteGraphIndices}
\bibfield{author}{\bibinfo{person}{Shulong Tan}, \bibinfo{person}{Weijie Zhao}, {and} \bibinfo{person}{Ping Li}.} \bibinfo{year}{2021}\natexlab{}.
\newblock \showarticletitle{Fast Neural Ranking on Bipartite Graph Indices}.
\newblock \bibinfo{journal}{\emph{Proc. {VLDB} Endow.}} \bibinfo{volume}{15}, \bibinfo{number}{4} (\bibinfo{year}{2021}), \bibinfo{pages}{794--803}.
\newblock
\urldef\tempurl%
\url{https://doi.org/10.14778/3503585.3503589}
\showDOI{\tempurl}


\bibitem[Tan et~al\mbox{.}(2020)]%
        {NeuralNetworkFastItemRanking}
\bibfield{author}{\bibinfo{person}{Shulong Tan}, \bibinfo{person}{Zhixin Zhou}, \bibinfo{person}{Zhaozhuo Xu}, {and} \bibinfo{person}{Ping Li}.} \bibinfo{year}{2020}\natexlab{}.
\newblock \showarticletitle{Fast Item Ranking under Neural Network based Measures}. In \bibinfo{booktitle}{\emph{{WSDM} '20: The Thirteenth {ACM} International Conference on Web Search and Data Mining, Houston, TX, USA, February 3-7, 2020}}, \bibfield{editor}{\bibinfo{person}{James Caverlee}, \bibinfo{person}{Xia~(Ben) Hu}, \bibinfo{person}{Mounia Lalmas}, {and} \bibinfo{person}{Wei Wang}} (Eds.). \bibinfo{publisher}{{ACM}}, \bibinfo{pages}{591--599}.
\newblock
\urldef\tempurl%
\url{https://doi.org/10.1145/3336191.3371830}
\showDOI{\tempurl}


\bibitem[Thakur et~al\mbox{.}(2021)]%
        {BEIR}
\bibfield{author}{\bibinfo{person}{Nandan Thakur}, \bibinfo{person}{Nils Reimers}, \bibinfo{person}{Andreas R{\"{u}}ckl{\'{e}}}, \bibinfo{person}{Abhishek Srivastava}, {and} \bibinfo{person}{Iryna Gurevych}.} \bibinfo{year}{2021}\natexlab{}.
\newblock \showarticletitle{{BEIR:} {A} Heterogeneous Benchmark for Zero-shot Evaluation of Information Retrieval Models}. In \bibinfo{booktitle}{\emph{Proceedings of the Neural Information Processing Systems Track on Datasets and Benchmarks 1, NeurIPS Datasets and Benchmarks 2021, December 2021, virtual}}, \bibfield{editor}{\bibinfo{person}{Joaquin Vanschoren} {and} \bibinfo{person}{Sai{-}Kit Yeung}} (Eds.).
\newblock
\urldef\tempurl%
\url{https://datasets-benchmarks-proceedings.neurips.cc/paper/2021/hash/65b9eea6e1cc6bb9f0cd2a47751a186f-Abstract-round2.html}
\showURL{%
\tempurl}


\bibitem[Vapnik and Chervonenkis(1971)]%
        {VCDim}
\bibfield{author}{\bibinfo{person}{V.~N. Vapnik} {and} \bibinfo{person}{A.~Ya. Chervonenkis}.} \bibinfo{year}{1971}\natexlab{}.
\newblock \showarticletitle{On the Uniform Convergence of Relative Frequencies of Events to Their Probabilities}.
\newblock \bibinfo{journal}{\emph{Theory of Probability \& Its Applications}} \bibinfo{volume}{16}, \bibinfo{number}{2} (\bibinfo{year}{1971}), \bibinfo{pages}{264--280}.
\newblock
\urldef\tempurl%
\url{https://doi.org/10.1137/1116025}
\showDOI{\tempurl}


\bibitem[Vaswani et~al\mbox{.}(2017)]%
        {AttentionIsAllYouNeed}
\bibfield{author}{\bibinfo{person}{Ashish Vaswani}, \bibinfo{person}{Noam Shazeer}, \bibinfo{person}{Niki Parmar}, \bibinfo{person}{Jakob Uszkoreit}, \bibinfo{person}{Llion Jones}, \bibinfo{person}{Aidan~N. Gomez}, \bibinfo{person}{Lukasz Kaiser}, {and} \bibinfo{person}{Illia Polosukhin}.} \bibinfo{year}{2017}\natexlab{}.
\newblock \showarticletitle{Attention is All you Need}. In \bibinfo{booktitle}{\emph{Advances in Neural Information Processing Systems 30: Annual Conference on Neural Information Processing Systems 2017, December 4-9, 2017, Long Beach, CA, {USA}}}, \bibfield{editor}{\bibinfo{person}{Isabelle Guyon}, \bibinfo{person}{Ulrike von Luxburg}, \bibinfo{person}{Samy Bengio}, \bibinfo{person}{Hanna~M. Wallach}, \bibinfo{person}{Rob Fergus}, \bibinfo{person}{S.~V.~N. Vishwanathan}, {and} \bibinfo{person}{Roman Garnett}} (Eds.). \bibinfo{pages}{5998--6008}.
\newblock
\urldef\tempurl%
\url{https://proceedings.neurips.cc/paper/2017/hash/3f5ee243547dee91fbd053c1c4a845aa-Abstract.html}
\showURL{%
\tempurl}


\bibitem[von Oswald et~al\mbox{.}(2020)]%
        {ContinualLearningWithHypernetworks}
\bibfield{author}{\bibinfo{person}{Johannes von Oswald}, \bibinfo{person}{Christian Henning}, \bibinfo{person}{Jo{\~{a}}o Sacramento}, {and} \bibinfo{person}{Benjamin~F. Grewe}.} \bibinfo{year}{2020}\natexlab{}.
\newblock \showarticletitle{Continual learning with hypernetworks}. In \bibinfo{booktitle}{\emph{8th International Conference on Learning Representations, {ICLR} 2020, Addis Ababa, Ethiopia, April 26-30, 2020}}. \bibinfo{publisher}{OpenReview.net}.
\newblock
\urldef\tempurl%
\url{https://openreview.net/forum?id=SJgwNerKvB}
\showURL{%
\tempurl}


\bibitem[Voorhees(2004)]%
        {Robust04}
\bibfield{author}{\bibinfo{person}{Ellen~M. Voorhees}.} \bibinfo{year}{2004}\natexlab{}.
\newblock \showarticletitle{Overview of the {TREC} 2004 Robust Track}. In \bibinfo{booktitle}{\emph{Proceedings of the Thirteenth Text REtrieval Conference, {TREC} 2004, Gaithersburg, Maryland, USA, November 16-19, 2004}} \emph{(\bibinfo{series}{{NIST} Special Publication}, Vol.~\bibinfo{volume}{500-261})}, \bibfield{editor}{\bibinfo{person}{Ellen~M. Voorhees} {and} \bibinfo{person}{Lori~P. Buckland}} (Eds.). \bibinfo{publisher}{National Institute of Standards and Technology {(NIST)}}.
\newblock
\urldef\tempurl%
\url{http://trec.nist.gov/pubs/trec13/papers/ROBUST.OVERVIEW.pdf}
\showURL{%
\tempurl}


\bibitem[Wang et~al\mbox{.}(2023)]%
        {SimLM}
\bibfield{author}{\bibinfo{person}{Liang Wang}, \bibinfo{person}{Nan Yang}, \bibinfo{person}{Xiaolong Huang}, \bibinfo{person}{Binxing Jiao}, \bibinfo{person}{Linjun Yang}, \bibinfo{person}{Daxin Jiang}, \bibinfo{person}{Rangan Majumder}, {and} \bibinfo{person}{Furu Wei}.} \bibinfo{year}{2023}\natexlab{}.
\newblock \showarticletitle{{S}im{LM}: Pre-training with Representation Bottleneck for Dense Passage Retrieval}. In \bibinfo{booktitle}{\emph{Proceedings of the 61st Annual Meeting of the Association for Computational Linguistics (Volume 1: Long Papers)}}, \bibfield{editor}{\bibinfo{person}{Anna Rogers}, \bibinfo{person}{Jordan Boyd-Graber}, {and} \bibinfo{person}{Naoaki Okazaki}} (Eds.). \bibinfo{publisher}{Association for Computational Linguistics}, \bibinfo{address}{Toronto, Canada}, \bibinfo{pages}{2244--2258}.
\newblock
\urldef\tempurl%
\url{https://doi.org/10.18653/v1/2023.acl-long.125}
\showDOI{\tempurl}


\bibitem[Weller et~al\mbox{.}(2024a)]%
        {FollowIR}
\bibfield{author}{\bibinfo{person}{Orion Weller}, \bibinfo{person}{Benjamin Chang}, \bibinfo{person}{Sean MacAvaney}, \bibinfo{person}{Kyle Lo}, \bibinfo{person}{Arman Cohan}, \bibinfo{person}{Benjamin~Van Durme}, \bibinfo{person}{Dawn~J. Lawrie}, {and} \bibinfo{person}{Luca Soldaini}.} \bibinfo{year}{2024}\natexlab{a}.
\newblock \showarticletitle{FollowIR: Evaluating and Teaching Information Retrieval Models to Follow Instructions}.
\newblock \bibinfo{journal}{\emph{CoRR}}  \bibinfo{volume}{abs/2403.15246} (\bibinfo{year}{2024}).
\newblock
\urldef\tempurl%
\url{https://doi.org/10.48550/ARXIV.2403.15246}
\showDOI{\tempurl}
\showeprint[arXiv]{2403.15246}


\bibitem[Weller et~al\mbox{.}(2024b)]%
        {MSMARCO-with-Instructions}
\bibfield{author}{\bibinfo{person}{Orion Weller}, \bibinfo{person}{Benjamin~Van Durme}, \bibinfo{person}{Dawn~J. Lawrie}, \bibinfo{person}{Ashwin Paranjape}, \bibinfo{person}{Yuhao Zhang}, {and} \bibinfo{person}{Jack Hessel}.} \bibinfo{year}{2024}\natexlab{b}.
\newblock \showarticletitle{Promptriever: Instruction-Trained Retrievers Can Be Prompted Like Language Models}.
\newblock \bibinfo{journal}{\emph{CoRR}}  \bibinfo{volume}{abs/2409.11136} (\bibinfo{year}{2024}).
\newblock
\urldef\tempurl%
\url{https://doi.org/10.48550/ARXIV.2409.11136}
\showDOI{\tempurl}
\showeprint[arXiv]{2409.11136}


\bibitem[Xiao et~al\mbox{.}(2022)]%
        {RetroMAE}
\bibfield{author}{\bibinfo{person}{Shitao Xiao}, \bibinfo{person}{Zheng Liu}, \bibinfo{person}{Yingxia Shao}, {and} \bibinfo{person}{Zhao Cao}.} \bibinfo{year}{2022}\natexlab{}.
\newblock \showarticletitle{RetroMAE: Pre-Training Retrieval-oriented Language Models Via Masked Auto-Encoder}. In \bibinfo{booktitle}{\emph{Proceedings of the 2022 Conference on Empirical Methods in Natural Language Processing, {EMNLP} 2022, Abu Dhabi, United Arab Emirates, December 7-11, 2022}}, \bibfield{editor}{\bibinfo{person}{Yoav Goldberg}, \bibinfo{person}{Zornitsa Kozareva}, {and} \bibinfo{person}{Yue Zhang}} (Eds.). \bibinfo{publisher}{Association for Computational Linguistics}, \bibinfo{pages}{538--548}.
\newblock
\urldef\tempurl%
\url{https://doi.org/10.18653/V1/2022.EMNLP-MAIN.35}
\showDOI{\tempurl}


\bibitem[Xiong et~al\mbox{.}(2017)]%
        {KNRM}
\bibfield{author}{\bibinfo{person}{Chenyan Xiong}, \bibinfo{person}{Zhuyun Dai}, \bibinfo{person}{Jamie Callan}, \bibinfo{person}{Zhiyuan Liu}, {and} \bibinfo{person}{Russell Power}.} \bibinfo{year}{2017}\natexlab{}.
\newblock \showarticletitle{End-to-End Neural Ad-hoc Ranking with Kernel Pooling}. In \bibinfo{booktitle}{\emph{Proceedings of the 40th International {ACM} {SIGIR} Conference on Research and Development in Information Retrieval, Shinjuku, Tokyo, Japan, August 7-11, 2017}}, \bibfield{editor}{\bibinfo{person}{Noriko Kando}, \bibinfo{person}{Tetsuya Sakai}, \bibinfo{person}{Hideo Joho}, \bibinfo{person}{Hang Li}, \bibinfo{person}{Arjen~P. de~Vries}, {and} \bibinfo{person}{Ryen~W. White}} (Eds.). \bibinfo{publisher}{{ACM}}, \bibinfo{pages}{55--64}.
\newblock
\urldef\tempurl%
\url{https://doi.org/10.1145/3077136.3080809}
\showDOI{\tempurl}


\bibitem[Xiong et~al\mbox{.}(2021)]%
        {ANCE}
\bibfield{author}{\bibinfo{person}{Lee Xiong}, \bibinfo{person}{Chenyan Xiong}, \bibinfo{person}{Ye Li}, \bibinfo{person}{Kwok{-}Fung Tang}, \bibinfo{person}{Jialin Liu}, \bibinfo{person}{Paul~N. Bennett}, \bibinfo{person}{Junaid Ahmed}, {and} \bibinfo{person}{Arnold Overwijk}.} \bibinfo{year}{2021}\natexlab{}.
\newblock \showarticletitle{Approximate Nearest Neighbor Negative Contrastive Learning for Dense Text Retrieval}. In \bibinfo{booktitle}{\emph{9th International Conference on Learning Representations, {ICLR} 2021, Virtual Event, Austria, May 3-7, 2021}}. \bibinfo{publisher}{OpenReview.net}.
\newblock
\urldef\tempurl%
\url{https://openreview.net/forum?id=zeFrfgyZln}
\showURL{%
\tempurl}


\bibitem[Yang et~al\mbox{.}(2018)]%
        {DRMM}
\bibfield{author}{\bibinfo{person}{Zhou Yang}, \bibinfo{person}{Qingfeng Lan}, \bibinfo{person}{Jiafeng Guo}, \bibinfo{person}{Yixing Fan}, \bibinfo{person}{Xiaofei Zhu}, \bibinfo{person}{Yanyan Lan}, \bibinfo{person}{Yue Wang}, {and} \bibinfo{person}{Xueqi Cheng}.} \bibinfo{year}{2018}\natexlab{}.
\newblock \showarticletitle{A Deep Top-K Relevance Matching Model for Ad-hoc Retrieval}. In \bibinfo{booktitle}{\emph{Information Retrieval - 24th China Conference, {CCIR} 2018, Guilin, China, September 27-29, 2018, Proceedings}} \emph{(\bibinfo{series}{Lecture Notes in Computer Science}, Vol.~\bibinfo{volume}{11168})}, \bibfield{editor}{\bibinfo{person}{Shichao Zhang}, \bibinfo{person}{Tie{-}Yan Liu}, \bibinfo{person}{Xianxian Li}, \bibinfo{person}{Jiafeng Guo}, {and} \bibinfo{person}{Chenliang Li}} (Eds.). \bibinfo{publisher}{Springer}, \bibinfo{pages}{16--27}.
\newblock
\urldef\tempurl%
\url{https://doi.org/10.1007/978-3-030-01012-6\_2}
\showDOI{\tempurl}


\bibitem[Zamani et~al\mbox{.}(2018)]%
        {SNRM}
\bibfield{author}{\bibinfo{person}{Hamed Zamani}, \bibinfo{person}{Mostafa Dehghani}, \bibinfo{person}{W.~Bruce Croft}, \bibinfo{person}{Erik~G. Learned{-}Miller}, {and} \bibinfo{person}{Jaap Kamps}.} \bibinfo{year}{2018}\natexlab{}.
\newblock \showarticletitle{From Neural Re-Ranking to Neural Ranking: Learning a Sparse Representation for Inverted Indexing}. In \bibinfo{booktitle}{\emph{Proceedings of the 27th {ACM} International Conference on Information and Knowledge Management, {CIKM} 2018, Torino, Italy, October 22-26, 2018}}, \bibfield{editor}{\bibinfo{person}{Alfredo Cuzzocrea}, \bibinfo{person}{James Allan}, \bibinfo{person}{Norman~W. Paton}, \bibinfo{person}{Divesh Srivastava}, \bibinfo{person}{Rakesh Agrawal}, \bibinfo{person}{Andrei~Z. Broder}, \bibinfo{person}{Mohammed~J. Zaki}, \bibinfo{person}{K.~Sel{\c{c}}uk Candan}, \bibinfo{person}{Alexandros Labrinidis}, \bibinfo{person}{Assaf Schuster}, {and} \bibinfo{person}{Haixun Wang}} (Eds.). \bibinfo{publisher}{{ACM}}, \bibinfo{pages}{497--506}.
\newblock
\urldef\tempurl%
\url{https://doi.org/10.1145/3269206.3271800}
\showDOI{\tempurl}


\bibitem[Zeng et~al\mbox{.}(2023)]%
        {uia}
\bibfield{author}{\bibinfo{person}{Hansi Zeng}, \bibinfo{person}{Surya Kallumadi}, \bibinfo{person}{Zaid Alibadi}, \bibinfo{person}{Rodrigo Nogueira}, {and} \bibinfo{person}{Hamed Zamani}.} \bibinfo{year}{2023}\natexlab{}.
\newblock \showarticletitle{A Personalized Dense Retrieval Framework for Unified Information Access}. In \bibinfo{booktitle}{\emph{Proceedings of the 46th International ACM SIGIR Conference on Research and Development in Information Retrieval}} (Taipei, Taiwan) \emph{(\bibinfo{series}{SIGIR '23})}. \bibinfo{publisher}{Association for Computing Machinery}, \bibinfo{address}{New York, NY, USA}, \bibinfo{pages}{121–130}.
\newblock
\showISBNx{9781450394086}
\urldef\tempurl%
\url{https://doi.org/10.1145/3539618.3591626}
\showDOI{\tempurl}


\bibitem[Zeng et~al\mbox{.}(2025)]%
        {lion}
\bibfield{author}{\bibinfo{person}{Hansi Zeng}, \bibinfo{person}{Julian Killingback}, {and} \bibinfo{person}{Hamed Zamani}.} \bibinfo{year}{2025}\natexlab{}.
\newblock \showarticletitle{Scaling Sparse and Dense Retrieval in Decoder-Only LLMs}.
\newblock \bibinfo{journal}{\emph{ArXiv}}  \bibinfo{volume}{abs/2502.15526} (\bibinfo{year}{2025}).
\newblock
\urldef\tempurl%
\url{https://api.semanticscholar.org/CorpusID:276558404}
\showURL{%
\tempurl}


\bibitem[Zeng et~al\mbox{.}(2022)]%
        {CL-DRD}
\bibfield{author}{\bibinfo{person}{Hansi Zeng}, \bibinfo{person}{Hamed Zamani}, {and} \bibinfo{person}{Vishwa Vinay}.} \bibinfo{year}{2022}\natexlab{}.
\newblock \showarticletitle{Curriculum Learning for Dense Retrieval Distillation}. In \bibinfo{booktitle}{\emph{{SIGIR} '22: The 45th International {ACM} {SIGIR} Conference on Research and Development in Information Retrieval, Madrid, Spain, July 11 - 15, 2022}}, \bibfield{editor}{\bibinfo{person}{Enrique Amig{\'{o}}}, \bibinfo{person}{Pablo Castells}, \bibinfo{person}{Julio Gonzalo}, \bibinfo{person}{Ben Carterette}, \bibinfo{person}{J.~Shane Culpepper}, {and} \bibinfo{person}{Gabriella Kazai}} (Eds.). \bibinfo{publisher}{{ACM}}, \bibinfo{pages}{1979--1983}.
\newblock
\urldef\tempurl%
\url{https://doi.org/10.1145/3477495.3531791}
\showDOI{\tempurl}


\bibitem[Zhan et~al\mbox{.}(2021)]%
        {adore}
\bibfield{author}{\bibinfo{person}{Jingtao Zhan}, \bibinfo{person}{Jiaxin Mao}, \bibinfo{person}{Yiqun Liu}, \bibinfo{person}{Jiafeng Guo}, \bibinfo{person}{M. Zhang}, {and} \bibinfo{person}{Shaoping Ma}.} \bibinfo{year}{2021}\natexlab{}.
\newblock \showarticletitle{Optimizing Dense Retrieval Model Training with Hard Negatives}.
\newblock \bibinfo{journal}{\emph{Proceedings of the 44th International ACM SIGIR Conference on Research and Development in Information Retrieval}} (\bibinfo{year}{2021}).
\newblock
\urldef\tempurl%
\url{https://api.semanticscholar.org/CorpusID:233289894}
\showURL{%
\tempurl}


\bibitem[Zhang et~al\mbox{.}(2019)]%
        {GraphHyperNetworksNeuralArchitectureSearch}
\bibfield{author}{\bibinfo{person}{Chris Zhang}, \bibinfo{person}{Mengye Ren}, {and} \bibinfo{person}{Raquel Urtasun}.} \bibinfo{year}{2019}\natexlab{}.
\newblock \showarticletitle{Graph HyperNetworks for Neural Architecture Search}. In \bibinfo{booktitle}{\emph{7th International Conference on Learning Representations, {ICLR} 2019, New Orleans, LA, USA, May 6-9, 2019}}. \bibinfo{publisher}{OpenReview.net}.
\newblock
\urldef\tempurl%
\url{https://openreview.net/forum?id=rkgW0oA9FX}
\showURL{%
\tempurl}


\bibitem[Zhao et~al\mbox{.}(2024)]%
        {GradientPruningTowardFastNeuralRanking}
\bibfield{author}{\bibinfo{person}{Weijie Zhao}, \bibinfo{person}{Shulong Tan}, {and} \bibinfo{person}{Ping Li}.} \bibinfo{year}{2024}\natexlab{}.
\newblock \showarticletitle{{GUITAR:} Gradient Pruning toward Fast Neural Ranking}. In \bibinfo{booktitle}{\emph{Proceedings of the 47th International {ACM} {SIGIR} Conference on Research and Development in Information Retrieval, {SIGIR} 2024, Washington DC, USA, July 14-18, 2024}}, \bibfield{editor}{\bibinfo{person}{Grace~Hui Yang}, \bibinfo{person}{Hongning Wang}, \bibinfo{person}{Sam Han}, \bibinfo{person}{Claudia Hauff}, \bibinfo{person}{Guido Zuccon}, {and} \bibinfo{person}{Yi~Zhang}} (Eds.). \bibinfo{publisher}{{ACM}}, \bibinfo{pages}{163--173}.
\newblock
\urldef\tempurl%
\url{https://doi.org/10.1145/3626772.3657728}
\showDOI{\tempurl}


\end{thebibliography}

\end{document}